\documentclass[aps,prd,10pt,twocolumn,superscriptaddress,floatfix,nofootinbib,showpacs,longbibliography]{revtex4-2}
\usepackage[utf8]{inputenc}  
\usepackage[T1]{fontenc}     
\usepackage[british]{babel}  
\usepackage[colorlinks=true, citecolor=blue, urlcolor=blue]{hyperref}  
\usepackage{graphicx} 
\usepackage[babel]{microtype}  
\usepackage{amsmath,amssymb,amsthm,bm,amsfonts,mathrsfs,bbm} 

\usepackage{xspace}  
\usepackage{pgf,tikz}
\usepackage{xcolor}
\usepackage{multirow}
\usepackage{array}
\usepackage{bigstrut}
\usepackage{braket}
\usepackage{color}
\usepackage{natbib}
\usepackage{multirow}
\usepackage{mathtools}
\usepackage[normalem]{ulem}
\usepackage{float}
\usepackage{xcolor,colortbl}
\usepackage{color}
\usepackage{subcaption}
\newcommand{\Tr}{\operatorname{Tr}}

\newcommand{\be}{\begin{equation}}
\newcommand{\ee}{\end{equation}}
\newcommand{\ba}{\begin{eqnarray}}
\newcommand{\ea}{\end{eqnarray}}
\newcommand{\ketbra}[2]{|#1\rangle \langle #2|}

\newtheorem{theorem}{Theorem}
\newtheorem{corollary}{Corollary}

\newtheorem{proposition}{Proposition}
\newtheorem{observation}{Observation}

\usepackage{mathtools}

\usepackage{enumitem}

\begin{document}

\title{Interference between lossy quantum evolutions activates information backflow}

\author{Sutapa Saha}
\email{sutapa.gate@gmail.com}    
\affiliation{QIC Group, Harish Chandra Research Institute, A CI of Homi Bhabha National Institute, Chhatnag Road, Jhunsi, Prayagraj 211019, India}

\author{Ujjwal Sen}
\email{ujjwal@hri.res.in}    
\affiliation{QIC Group, Harish Chandra Research Institute, A CI of Homi Bhabha National Institute, Chhatnag Road, Jhunsi, Prayagraj 211019, India}

\begin{abstract}
Quantum evolutions are often non-unitary and in such cases, they are frequently 
regarded as lossy.
Such lossiness, however, does not necessarily persist throughout the evolution, and there can often be intermediate time-spans 
during which information ebbs in the environment to re-flood the system -- an event known as \textit{information backflow} (IB). This phenomenon serves as a well-established and sufficient indicator of non-Markovian behavior of open quantum dynamics. Nevertheless, not all non-Markovian dynamics exhibit such backflow. We find that when interference is allowed between two quantum evolutions that  individually generate non-Markovianity and yet do not exhibit IB, it becomes possible to retrieve information from the environment. Furthermore, we show that this setup  involving coherently-controlled quantum operation trajectories provides enhanced performance and is more robust compared to an alternate coherently-controlled arrangement of the quantum-switch.

\end{abstract}

\maketitle

\section{Introduction}

The presence of coherence in quantum systems, distinguishes the theory fundamentally from its classical counterpart. It underlies many of the advantages that quantum systems offer in information processing tasks such as communication \cite{ambainis2008,farkas2025}, computation \cite{deutsch1992,shor1994}, 
and cryptography \cite{bennett1984}. However, quantum coherence -- essentially the principle of superposition -- is not only limited to the static entities of the quantum formalism and setups, but also spreads over to their 
dynamical aspects~\cite{aharonov1990,oi2003,aberg2004}. The advantage of superposing multiple quantum trajectories in context of noise mitigation, potentially leveraging the key distribution and information processing tasks was first exemplified in \cite{gisin2005}. Important further works 
in this direction includes identification of similar utilities of superposition in 
classical as well as quantum communication~\cite{abbott2020,chiribella2019,vanrietvelde2021}, quantum computation~\cite{thompson2018}, and  quantum metrology~\cite{blondeau2021}. 
Dynamical superposition has also been taken over to the case of superposing different orders of multiple quantum operations,
namely the quantum-switch. Such a toy-model, in the goal of exploring a new causal structure, predicted by the structure of quantum gravity \cite{hardy2007} is first encountered in \cite{chiribella2013}.
Consequently, throughout the last decade, the significance of quantum-switch has also been demonstrated in various directions, viz., classical and quantum communication \cite{ebler2018,chiribella2021,chiribella2021a,bhattacharya2021,sazim2021,guha2023}, thermodynamics \cite{felce2020,guha2020,simonov2022,simonov2025}, metrology \cite{zhao2020}, both theoretically and experimentally \cite{rubino2017,goswami2018,guo2020}.

We switch gears. All deterministic quantum operations are unitary, but one can bring in auxiliary systems (often called environments) before applying a \emph{joint} unitary on the combined system-auxiliary duo. The system in that case undergoes a 
completely-positive trace-preserving (CPTP) map.
Although the information encoded in a quantum system may partially leak into the adjoint environment after the complete action of the CPTP map, this does not guarantee that in each of the individual time-slots, the map acts as a lossy dynamics. In fact, there are instances of CPTP maps for which in the intermediate time-scale, information ``flows back'' to the system from the environment. The phenomenon, often referred to as IB~\cite{breuer2009}, serves as a hallmark of non-Markovian dynamics \cite{breuer2002}. Significant attention has recently 
been devoted to research on 
non-Markovianity and IB, due to its successful applications to quantum technologies, including in quantum communication \cite{laine2014, bylicka2014}, entanglement distribution \cite{xiang2014}, thermodynamic work extraction \cite{thomas2108}, entanglement generation \cite{mirkin2019}, and quantum control \cite{reich2015}. 

We may now ask whether there is a  natural connection between the concept of dynamical superposition and the notion of IB. We remember here that relations between quantum coherence and IB are recognized in the literature within non-Markovian operations~\cite{benabdallah2025,hadipour2025,poggi2017,man2018,khurana2019,jahromi2019,wu2020,giorgi2020,paulson2022,benatti2024,liu2016}.
Also, IB is known to be activable by using a quantum-switch~\cite{maity2024} as well as an optical quantum-switch. The later one has been experimentally realized and its non-Markovianity has been characterized by measuring the IB from the corresponding heat baths to the system. \cite{tang2024}.
%
%

We note however that experiments simulating quantum-switches have recently encountered questions regarding  actual realization of the indefinite causal structure \cite{vilasini2024,vilasini2024a}. In contrast, the experimental realization of superposing multiple quantum trajectories -- though dependent on access to extended channel configurations -- are arguably simpler to realize experimentally 
\cite{lamoureux2005,lai2024,rubino2021}.
Keeping this in mind, here we explore the possibility of activation of IB, limiting ourselves to 
interfering quantum operation trajectories. Importantly, this interference 
does not necessitate consideration of an indefinite causal structure. In fact, we show that a simple interference setup of multiple information-losing trajectories  leads to IB for a wider range of quantum states than that of their switch simulation. 
We subsequently investigate the robustness of the quantum system that coherently controls the choice of multiple trajectories and find that its advantages -- both in terms of robustness and in terms of control system parameters -- surpasses those of the corresponding switch configuration. This approach therefore also allows for improved experimental visibility of the effect of IB activation, in comparison to the switch-dependent setup. 


\section{Markovian and non-markovian dynamics}

In the study of open quantum systems, a central concern is how a system interacts with its surrounding environment. Such interactions typically lead to decoherence and loss of quantum information. Assuming time-independent or memoryless system–environment interactions leads to Markovian dynamics, typically described by the Lindblad master equation \cite{gorini1978,lindblad1976}. However, such assumptions often fail to capture the complexity of realistic physical scenarios, where memory effects become significant, giving rise to non-Markovian evolution.

\subsection{Canonical form of master equation}
The evolution of a density matrix is generally described by a completely positive and trace-preserving (CPTP) linear map \cite{heinosaari2011,nielsen2000}, that is, $\rho(t)=\phi_t[\rho(0)]$; here $\rho(0)$ denotes the initial density matrix and $\rho(t)$ is the evolved state under the channel action $\phi_t$. Under quite general assumptions, the dynamics of the system’s density operator $\rho(t)$ is governed by a master equation of the form, 
\begin{align}\label{mem-ker}
\dot{\rho}(t) = -i[H_s,\rho(t)]+\int_0^t ds~ \mathcal{K}_{s,t}[\rho(s)],
\end{align}
where $H_S$ is the system Hamiltonian and $\mathcal{K}_{s,t}$, termed as the memory kernel, is a linear map that describes the effects of the environment on the system. 

Consider a CP map $\phi_t$ such that $\phi_t[\rho(0)]=\rho(t)$. Suppose this evolution process satisfies a memory-kernel master equation of form in Eq. (\ref{mem-ker}). If $\phi_t$ is assumed to be invertible for the time interval considered, \textit{i.e.}, $\phi_t^{-1}$ exists satisfying $\phi_t^{-1}\circ\phi_t=\mathbb{I}$, where $\mathbb{I}$ is the identity map and $\mathbb{I}(\rho)=\rho,~\forall\rho$, then Eq. (\ref{mem-ker}) can be written by absorbing the Hamiltonian component into $\mathcal{K}_{s,t}$ as
\begin{align}\label{time_loc}
    \dot{\rho}(t)= \int_0^t ds~ \mathcal{K}_{s,t}[\rho(s)]
    =\Lambda_t[\rho(t)],
\end{align}
where $\Lambda_t=\int_0^t ds~ (\mathcal{K}_{s,t}~\circ \phi_s\circ\phi_t^{-1})$ and corresponds to a linear map that is Hermiticity preserving and $\Lambda_t[\rho]$ is traceless for all $\rho$. Eq. (\ref{time_loc}) is referred to as time-local form of master equation \cite{hall2014,andersson2007}. Interestingly, any time-local master equation for a quantum system having a $d-$dimensional
Hilbert space, can be written in the canonical form \cite{hall2014}
\begin{align}\label{can_g}
 \dot{\rho}(t)= & -i[H_s,\rho(t)]\nonumber\\
 &+\sum_{i=1}^{d^2-1} \gamma_i(t)\,\Bigl(A_i(t)\,\rho\,A_i^\dagger(t)\;-\; \frac{1}{2}\,\{A_i^\dagger(t\,) A_i(t),\,\rho\}\Bigr)   
\end{align}
where the set $\{A_i(t)\}$ forms an orthonormal basis set of traceless operators and $\{\gamma_i(t)\}$ denote the decoherence rates.

Eq. (\ref{can_g}) closely resembles the Lindblad form typically associated with memoryless master equation \cite{breuer2002} which corresponds to the \emph{Markovian} dynamics. However, Eq. (\ref{can_g}) can be properly regarded as representing \textit{Markovian} dynamics, \textit{if and only if} all the decoherence rates remains positive over all times, that is $\gamma_i(t)\geq 0~\forall i,t$ \cite{wolf2008,breuer2009,rivas2010,hall2008}. Furthermore, the non-negativity of the decoherence rates $\gamma_i(t)$ guarantees that the corresponding dynamical evolution can be decomposed into a sequence of intermediate completely positive (CP) maps, i.e., the evolution is CP-divisible. Conversely, if any of the coefficients $\gamma_i$ assume negative values, the dynamics exhibit \textit{non-Markovian} behavior.

\subsection{Information Backflow}
The Markovian approximation, valid under weak or singular coupling to a memoryless reservoir, fails to capture memory effects and any influence the environment may exert back on the system. As a result, Markovian dynamics are characterized by a continuous and monotonic loss of distinguishability between quantum states over time. However, in realistic settings, the dynamics often depart from such idealized description \cite{breuer2002}. In many physical scenarios, information is not irreversibly lost to the environment; instead, it can temporarily flow back into the system. Consequently, during that intervals of the overall evolution, the distinguishability between quantum states may increase rather than decrease. This phenomenon, known as IB \cite{breuer2009}, signals a temporary reversal in the direction of information flow and marks a fundamental signature of non-Markovian behavior.

The distinguishability of two quantum states $\rho_1$ and $\rho_2$ can be measured in terms of the trace distance defined as 
\begin{align*}
    \mathcal{D}[\rho_1,\rho_2]=\frac{1}{2}\Tr\|\rho_1-\rho_2\|_1,
\end{align*}
where $\|A\|_1=\Tr[\sqrt{A^{\dagger}A}]$ represents the trace norm of an operator $A$. A remarkable feature of the trace distance is that it is monotonic with respect to the action of any CPTP map $\phi$,
\begin{align*}
    \mathcal{D}[\phi(\rho_1),\phi(\rho_2)]\leq\mathcal{D}[\rho_1,\rho_2],
\end{align*}
which readily implies
\begin{align}\label{blp}
    \frac{d}{dt}\Big[\mathcal{D}[\phi(\rho_1),\phi(\rho_2)]\Big]\leq0,
\end{align}

In addition, Eq.(\ref{blp}) is also satisfied by those dynamics that admit a decomposition into a continuous sequence of intermediate positive maps and can consequently be classified as P-divisible \cite{chruscinski2014}. However, in certain instances of \textit{non-Markovian} dynamics where CP-divisibility as well as P-divisibility does not hold, Eq. (\ref{blp}) is violated \cite{breuer2009}.  Physically, such behavior can be interpreted as the information flow in the reverse direction 
and thus resulting in a temporary increase in the distinguishability between two quantum states.

\section{coherent control of quantum 
channels}\label{config_cont}

In this section, we provide a concise overview of the mathematical framework describing the two forms of coherent control over quantum channel configurations: control over the choice of quantum operations and control over their causal ordering.

A convenient realization of quantum channels $\mathcal{E}$ can be expressed in terms of Kraus representation as follows:
\begin{align*}
    \mathcal{E}(\rho) = \sum_i E_i \rho E_i^\dagger,
\end{align*}
where the Kraus operators $\{E_i\}$ satisfy the normalization condition
\begin{align*}
    \sum_i E_i^\dagger E_i = \mathbb{I},
\end{align*}
with $\mathbb{I}$ denoting the identity operator on the system's Hilbert space.

\begin{figure}[htbp]
  \centering

  \begin{subfigure}[b]{0.23\textwidth}
    \includegraphics[width=1.1\linewidth]{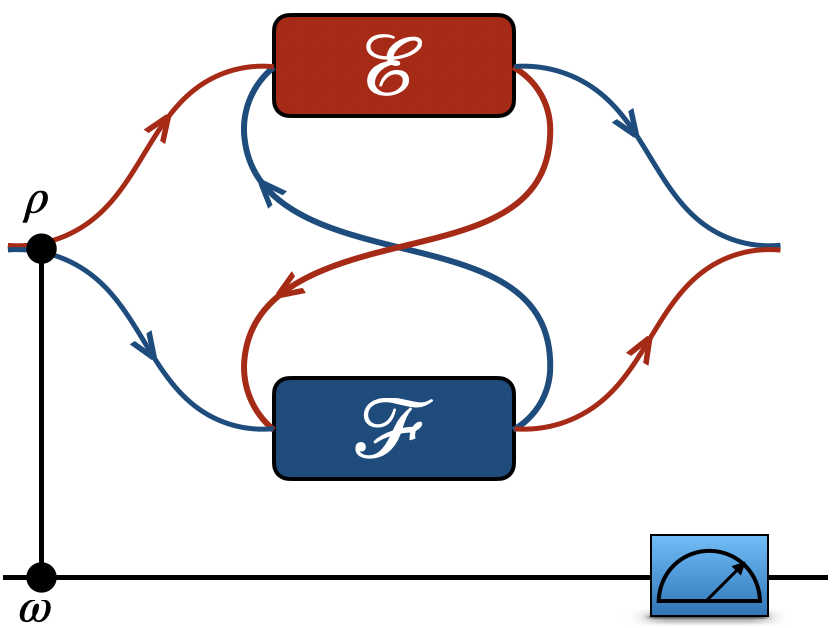}
    \caption{Coherent control over causal orders}
    \label{fig:a}
  \end{subfigure}
  \hfill
  \begin{subfigure}[b]{0.23\textwidth}
    \includegraphics[width=1.1\linewidth]{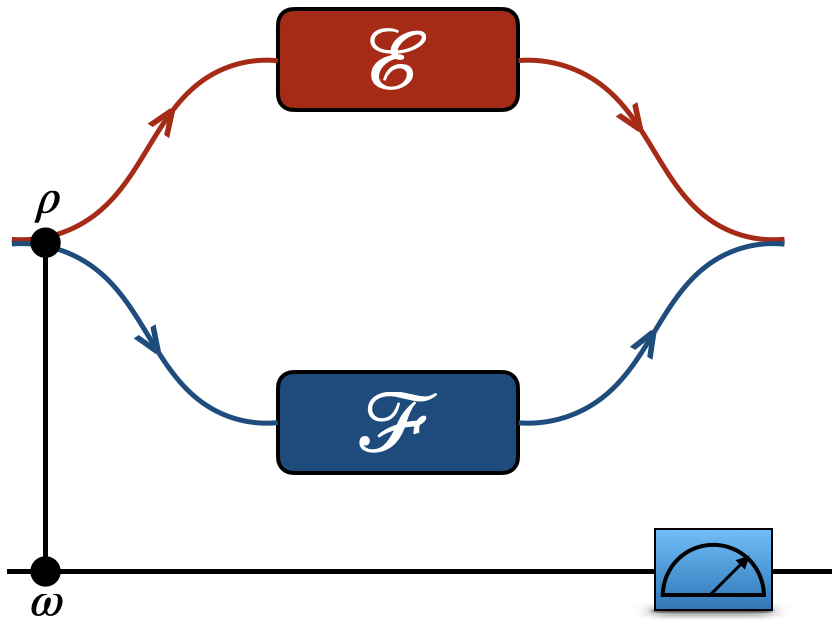}
    \caption{Coherent control over the choice of the channels}
    \label{fig:b}
  \end{subfigure}

  \caption{Depiction of the two paradigmatic coherent control configurations. We depict here the two paradigmatic coherent control configurations of quantum channels, viz. (a) the coherent control over causal orders and (b) the coherent control over quantum operation trajectories.}
  \label{combined}
\end{figure}

The coherent control over the ordering of two quantum channels is commonly captured within the framework of the quantum-switch \cite{chiribella2013}. Mathematically, the quantum-switch defines a higher order operation where two channels $\mathcal{E}$ and $\mathcal{F}$ acting on a system $\rho$, referred to as the target, and produces a new channel $\mathcal{S}(\mathcal{E},\mathcal{F})$. This resulting channel acts jointly on both the target system and an auxiliary \textit{control} system $\omega$. Under the switch action $\mathcal{S}(\mathcal{E}, \mathcal{F}),~\mathcal{E}$ and $\mathcal{F}$ are applied either in the order of  $\mathcal{E} \circ \mathcal{F}$ or in the order of $\mathcal{F} \circ \mathcal{E}$, depending on whether $\omega$ is prepared in the state $\ket{0}$ or $\ket{1}$, respectively.

Hence, the output of the quantum-switch can be expressed as,
\begin{align*}
   \mathcal{S}_{\mathcal{E}, \mathcal{F}}(\rho\otimes\omega) = \sum_{i,j} S_{ij}\, (\rho\otimes\omega)\, S_{ij}^\dagger, 
\end{align*}
where the Kraus operators $\{S_{ij}\}$ of $\mathcal{S}(\mathcal{E}, \mathcal{F})$ are given as
\begin{align*}
    S_{ij}=E_iF_j\otimes\ketbra{0}{0}+F_jE_i\otimes\ketbra{1}{1},
\end{align*}
with $\{E_i\}$ and $\{F_j\}$ denoting the Kraus operators of the channels $\mathcal{E}$ and $\mathcal{F}$ respectively.

In contrast, coherent control over the choice of two channels involves a control-target system in which the control system determines which channel is applied to transmit the quantum state, rather than superposing the order in which two channels are applied \cite{abbott2020,aberg2004,chiribella2019}. In this case, the quantum channel $\mathcal{N}$ that defines the path of evolution of $\rho$ in either channel $\mathcal{E}$ or channel $\mathcal{F}$ depending of the state of a control system has the form 
\begin{align*}
    \mathcal{N}_{\mathcal{E}, \mathcal{F}}(\rho\otimes\omega)=\sum_{ij}N_{ij}\, (\rho\otimes\omega)\, N_{ij}^\dagger, 
\end{align*}
with 
\begin{align*}
    N_{ij}=\alpha_j\,E_i\otimes\ketbra{0}{0}+\beta_i\,F_j\otimes\ketbra{1}{1},
\end{align*}
where $\alpha_j$ and $\beta_i$ are complex amplitudes and satisfy the normalization condition $\sum_j|\alpha_j|^2=\sum_i|\beta_i|^2=1$.

Note that while the supermap describing quantum-switch, \textit{i.e.} the control-order channel $\mathcal{S}(\mathcal{E}, \mathcal{F})$ depends solely on the quantum channels $\mathcal{E}$ and $\mathcal{F}$, the control-choice channel $\mathcal{N}$ depends additionally on the complex amplitudes  $\{\alpha_i\}$ and $\{\beta_j\}$ \cite{oi2003, aberg2004, chiribella2019, abbott2020, vanrietvelde2021,guha2023}. This dependence arises from the specific mechanism by which the control system determines which channel is effectively applied to the target system. A detailed physical interpretation of this dependence is provided in terms of the \textit{vacuum extension} $\widetilde{\mathcal{E}}$ of a quantum channel $\mathcal{E}$ as discussed in Ref. \cite{chiribella2019}. In this construction, the action of a channel is extended to include the possibility that the system is not transmitted through it, corresponding effectively to feeding the channel a trivial input state such as the vacuum.

Formally, the Kraus operators of the extended channels $\widetilde{\mathcal{E}}$ and $\widetilde{\mathcal{F}}$ take the form
\begin{align*}
 \widetilde{E}_i = E_i + \beta_i \ketbra{\text{vac}}{\text{vac}}, \quad \widetilde{F}_j = F_j + \alpha_j \ketbra{\text{vac}}{\text{vac}},   
\end{align*}
respectively, where $\ket{\text{vac}}$ denotes the vacuum state orthogonal to all states of the target Hilbert space \cite{chiribella2019}.

\section{Activation of information backflow}

Let us consider the time-local qubit master equation
\begin{equation}\label{ete_nm}
\dot{\rho}(t)=\sum_{i=1}^3 \frac{\gamma_i}{2}[\sigma_i\,\rho(t)\,\sigma_i-\rho(t)]
\end{equation}
where $\sigma_i$ are the pauli matrices and $\gamma_i$ can be in general $t$ dependent. For our specific interest, we choose $\gamma_1=\gamma_2=1$ and $\gamma_3=-\tanh{t}$ \cite{hall2014}. In this context, negative value of $\gamma_3$ carries a clear signature of non-Markovian behavior. Furthermore, since $\gamma_3$ remains negative throughout the entire evolution, the process exhibits \textit{eternal non-Markovian} dynamics \cite{vaishy2022,hall2014,utagi2024,dąbrowska2021}. 

Following the procedure described in Refs. \cite{hall2014,andersson2007}, the Kraus operators of the corresponding dynamical map $\phi_t$ \cite{maity2024} can be expressed as follows,
\begin{equation}\label{kraus}
\begin{aligned}
K_1(t) &= \sqrt{q_2(t)} 
\begin{bmatrix}
0 & 1 \\
0 & 0
\end{bmatrix}
\\[8pt]
K_2(t) &= \sqrt{q_2(t)} 
\begin{bmatrix}
0 & 0 \\
1 & 0
\end{bmatrix}
\\[8pt]
K_3(t) &= \sqrt{\frac{q_1(t) + q_3(t)}{2}}\,
\begin{bmatrix}
1 & 0 \\
0 & 1
\end{bmatrix}
\\[8pt]
K_4(t) &= \sqrt{\frac{q_1(t)-q_3(t)}{2}}\,
\begin{bmatrix}
-1 & 0 \\
0 & 1
\end{bmatrix}
\end{aligned}
\end{equation}
where, $q_1(t)=q_3(t)=\frac{1+e^{-2t}}{2}$ and $q_2(t)=\frac{1-e^{-2t}}{2}$. Evidently, $K_4=0$ and hence the evolution of any density matrix $\rho(0)$, under the channel action of $\phi_t$, is given as $\rho(t)=\phi_t[\rho(0)]=\sum_{i=1}^3K_i\rho(0)K_i^{\dagger}$. 

Before moving further with coherent configuration of this quantum channel $\phi_t$, we would like to highlight a crucial feature of it.
\begin{observation}
    The dynamics described by Eq. (\ref{ete_nm}) is P-divisible and hence admits no IB.
\end{observation}
\begin{proof}
A dynamics of the form given in Eq. (\ref{ete_nm}) is P-divisible \textit{if and only if} $\gamma_i+\gamma_j\geq 0$ \cite{chruscinski2014}. In our case, $\gamma_i+\gamma_j$ equal to either $2$ or $1-\tanh t$. Both of these quantities are positive for all values of $t\geq 0$ which immediately confirms the P-divisibility of the given dynamics.

As a consequence, IB can never be observed in this dynamical evolution. This, in turn, implies that the dynamics remains P-divisible throughout the entire evolution under consideration.
\end{proof}

Although the phenomenon of IB is absent in the present dynamics, it is an important question whether it can be triggered by employing the same dynamical map in some combined form. More specifically, one may ask whether IB could emerge if the map were used in a concatenated configuration, either in series or in parallel. In the appendix of Ref. \cite{maity2024}, the authors have addressed this question and concluded that, under a definite causal structure, the answer is negative. Furthermore, it has been proved that even when two copies of the dynamical map $\phi_t$ are used and a convex combination of their distinct causal orderings is taken, the dynamics remain devoid of IB.

Naturally, the next possibility that arises is whether such activation could be achieved through a coherently controlled configuration. In the following, we consider both types of coherently controlled configurations of quantum channels as discussed in Section~\ref{config_cont}. To this end, we take two copies of the channel $\phi_t$ together with an additional control system prepared in the state $\omega$. The final state $\rho(t)$ can be obtained after measuring the control qubit $\omega$ in some coherent basis. Depending on the preparation of $\omega$, the following cases arise.

\subsection{Maximally coherently-controlled configuration}

In this setting, the control system $\omega$ is initialized in the state $\ketbra{+}{+}$, with $\ket{+}=\tfrac{1}{\sqrt{2}}(\ket{0}+\ket{1})$. The target system, on the other hand, is assumed to start in the general state  $\rho(0)$.

We begin by analyzing the configuration wherein the control system governs the selection of the path traversed by the target system. Here, without loss of generality, we choose the values of the complex amplitudes as $\alpha_1=\alpha_2=\beta_1=\beta_2=\frac{1}{\sqrt{2}}$ and $\alpha_3=\beta_3=0$. Under this configuration, the Kraus operators $\{N_{ij}\}$ are given as, $N_{ij}=\alpha_jK_i\otimes\ketbra{0}{0}+\beta_iK_j\otimes\ketbra{1}{1}$ where $\{K_i\}$ are as given in Eq. (\ref{kraus}).
Hence, after carrying out the measurement in $\{\ketbra{\pm}{\pm}\}$ basis on the control system, the final state of the target $\rho(t)$ corresponding to the $\ketbra{+}{+}$ outcome of the controller can be written as,
    \begin{align}\label{max_co_eq_ps}
    \rho^I(t)=\frac{\bra{+}\left(\sum_{i,j=1}^3N_{ij}[\rho(0)\otimes\omega]N_{ij}^{\dagger}\right)\ket{+}}{\Tr\left[\bra{+}\left(\sum_{i,j=1}^3N_{ij}[\rho(0)\otimes\omega]N_{ij}^{\dagger}\right)\ket{+}\right]}\nonumber\\[2ex]
\end{align}

To check whether this configuration can activate IB or not, let us first consider the following two states as initials,
\begin{align}\label{tr.dist.}
 \rho_1(0)=&\frac{1}{2}(\mathbb{I}+\vec{r}.\vec{\sigma})\nonumber\\
 \rho_2(0)=&\frac{1}{2}(\mathbb{I}+\vec{s}.\vec{\sigma}),
\end{align}
where, $\vec{r}$ and $\vec{s}$ are the Bloch vectors of $\rho_1(0)$ and $\rho_2(0)$. The corresponding evolved state according to Eq. (\ref{max_co_eq_ps}) can be written in terms of their new Bloch vectors $\vec{r}^I,\vec{s}^I$ as,
\begin{align}\label{tr.dist.ps}
 \rho_1^I(t)=&\frac{1}{2}(\mathbb{I}+\vec{r}^I.\vec{\sigma})\nonumber\\
 \rho_2^I(t)=&\frac{1}{2}(\mathbb{I}+\vec{s}^I.\vec{\sigma}).   
\end{align}
A straightforward calculation leads us to write the new Bloch vectors in terms of the old ones as $\{\vec{r}^I,\vec{s}^I\}=M\{\vec{r},\vec{s}\}+\vec{c}$. Here, $M$ is a $3\times 3$ real matrix and $\vec{c}\in\mathbb{R}^3$. Under the quantum process described in Eq. (\ref{max_co_eq_ps}), we obtain $\vec{c}=\vec{0}$ and  
    \begin{align*}
        M=\frac{1}{5e^{2t}-1}\begin{pmatrix}
            3e^{2t}+1 & 0 & 0\\
            0 & e^{2t}+3 & 0\\
            0 & 0 & -e^{2t}+5
        \end{pmatrix}
    \end{align*}
Hence, the trace distance between the states of Eq. (\ref{tr.dist.ps}) under the interference of the channel $\phi_t$ can be written as
\begin{align}\label{e_norm_ps}
   \mathcal{D}[\rho_1^I(t),\rho_2^I(t)]=&\frac{1}{2}\|\vec{r}^I-\vec{s}^I\|\nonumber\\
   =&\frac{1}{2}\Big[M_{11}^2(r_1-s_1)^2+M_{22}^2(r_2-s_2)^2\nonumber\\
   &+M_{33}^2(r_3-s_3)^2\Big]^{\frac{1}{2}}
\end{align}
Here, $\|.\|$ denotes the Eucledian norm in $\mathbb{R}^3$. The time derivative of the trace distance given in  Eq. (\ref{e_norm_ps}) can be written as
\begin{align*}
  \frac{d}{dt}\mathcal{D}[\rho_1^I(t),\rho_2^I(t)]=-\frac{\xi^I(r_i,s_i,t)}{\mathcal{D}[\rho_1^I(t),\rho_2^I(t)]}
\end{align*}
where
\begin{align*}
  \xi^I(r_i,s_i,t)=&\frac{4e^{2t}}{(5e^{2t}-1)^2}\Big[M_{11}(r_1-s_1)^2\\
    &+2M_{22}(r_2-s_2)^2+3M_{33}(r_3-s_3)^2\Big]
\end{align*}
Now, we are led to the following proposition.
\begin{proposition}\label{prop_max_co_ps}
    The pair of post-measurement states, given in Eq. (\ref{tr.dist.ps}), will exhibit IB at an instant $t$ if and only if
\begin{align*}
    (i)&~t>\frac{1}{2}\ln 5\\
    (ii)&~(r_3-s_3)^2>\frac{M_{11}(r_1-s_1)^2
    +2M_{22}(r_2-s_2)^2}{3|M_{33}|}
\end{align*}
\end{proposition}
\begin{proof}
    IB will occur under the interfering configuration of the channel $\phi_t$ \textit{if and only if} $\frac{d}{dt}\mathcal{D}[\rho_1^I(t),\rho_2^I(t)]>0$, which readily implies
    \begin{align}\label{IB_ps}
      \implies  M_{11}(r_1-s_1)^2+&2M_{22}(r_2-s_2)^2\nonumber\\
      &+3M_{33}(r_3-s_3)^2<0
    \end{align}
     However, the above inequality holds \textit{only if} $M_{33}$ is strictly negative as all the terms in the above expression are positive except it. Thus, 
    \begin{align*}
        &M_{33}<0
        \implies  t>\frac{1}{2}\ln 5
    \end{align*}
    This proves the \textit{necessary} condition.\\
    Since, achieving IB requires $M_{33}=-|M_{33}|$, the inequality (\ref{IB_ps}) will be satisfied if
    \begin{align*}
        -3|M_{33}|(r_3-s_3)^2<&
        -[M_{11}(r_1-s_1)^2\\
        &+2M_{22}(r_2-s_2)^2]\\
        \implies  (r_3-s_3)^2>&\frac{M_{11}(r_1-s_1)^2
    +2M_{22}(r_2-s_2)^2}{3|M_{33}|}
    \end{align*}
    This completes the proof.
    \end{proof}
    Consider the initial Bloch vectors of the states given in Eq. (\ref{tr.dist.}) as 
    \begin{align}\label{same_azi}
        r_1&=\sin\theta_1\cos\Phi ~~&~~ s_1&=\sin\theta_2\cos\Phi\nonumber\\
        r_2&=\sin\theta_1\sin\Phi ~~&~~ s_2&=\sin\theta_2\sin\Phi\nonumber\\
        r_3&=\cos\theta_1 ~~&~~ s_3&=\cos\theta_2
    \end{align}
    Such sets of Bloch vectors leads us to the following corollary
    :
\begin{corollary}\label{cor_max_co_ps}
    If the Bloch vectors of the initial states consider same $\Phi$, IB occurs when
    \begin{align*}
        \arccos\left[\frac{e^{2t}-37-(e^{2t}-5)\cos 2\Phi}{11e^{2t}-23+(e^{2t}-5)\cos 2\Phi}\right]<(\theta_{1}+\theta_{2})\\<2\pi-\arccos\left[\frac{e^{2t}-37-(e^{2t}-5)\cos 2\Phi}{11e^{2t}-23+(e^{2t}-5)\cos 2\Phi}\right]
    \end{align*}
    where $n\in\mathbb{Z}_0^+$.
\end{corollary}

Now, consider the dynamics under the configuration of quantum-switch, \textit{i.e.} coherently controlled configuration where the causal orders of two quantum operations are kept into superposition. In our context, the Kraus operators of such configuration would be $S_{ij}=K_iK_j\otimes\ketbra{0}{0}+K_jK_i\otimes\ketbra{1}{1}$ and the state of the target $\rho(t)$ corresponding to the $\ketbra{+}{+}$ outcome of the control system would be given by,
\begin{align*}
\rho^S(t)&=\frac{\bra{+}\left(\sum_{i,j=1}^3S_{ij}[\rho(0)\otimes\omega]S_{ij}^{\dagger}\right)\ket{+}}{\Tr\left[\bra{+}\left(\sum_{i,j=1}^3S_{ij}[\rho(0)\otimes\omega]S_{ij}^{\dagger}\right)\ket{+}\right]}
    \end{align*}

Hence, the pair of states given in Eq. (\ref{tr.dist.}) has turned into the following outputs,
\begin{align}\label{tr.dist.qs}
 \rho_1^S(t)=&\frac{1}{2}(\mathbb{I}+\vec{r}^S.\vec{\sigma})\nonumber\\
 \rho_2^S(t)=&\frac{1}{2}(\mathbb{I}+\vec{s}^S.\vec{\sigma}).   
\end{align}
Here, the evolved Bloch vectors are given as $\{\vec{r^S},\vec{s}^S\}=N\{\vec{r},\vec{s\}}+\vec{d}$, where $N$ represents a $3\times 3$ real matrix and $\vec{d}$ is a vector in $\mathbb{R}^3$. Following the same procedure outlined above, we obtain $\vec{d}=\vec{0}$ and 
\begin{align*}
        &N_{ij}=0~\forall i\neq j;\\
        &N_{11}=N_{22}=\frac{3e^{4t}+2e^{2t}+3}{7e^{4t}+2e^{2t}-1},~N_{33}=-\frac{e^{4t}-2e^{2t}-7}{7e^{4t}+2e^{2t}-1}.
\end{align*}
Note that the evolution induced by the quantum-switch configuration preserves rotational symmetry about the $z$-axis.
\begin{figure*}[htbp]
  \centering
  \begin{subfigure}[b]{0.48\textwidth}
    \includegraphics[width=\linewidth]{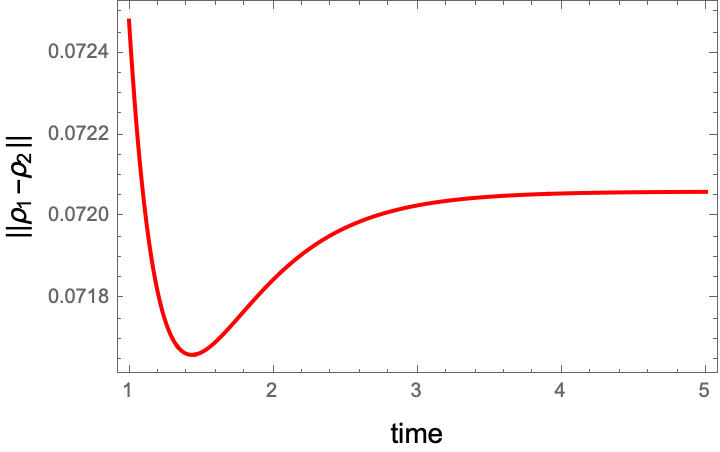}
    \caption{}
    \label{fig:a}
  \end{subfigure}
  \hfill
  \begin{subfigure}[b]{0.48\textwidth}
    \includegraphics[width=\linewidth]{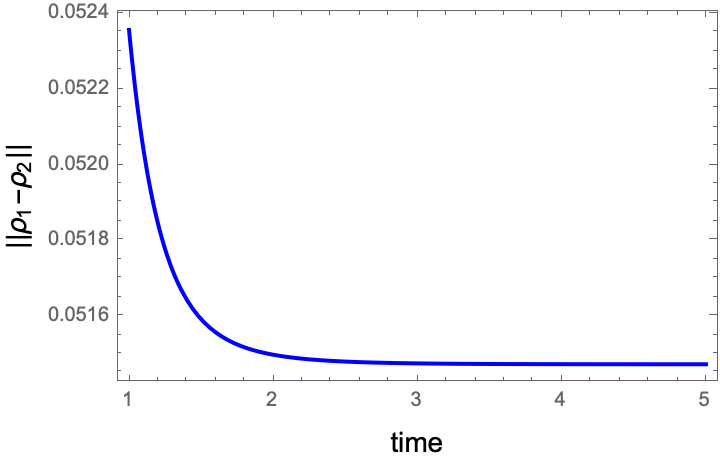}
    \caption{}
    \label{fig:b}
  \end{subfigure}
  \caption{IB or not: Time-dynamics of the distance between two states for superposing trajectories and for quantum-switch. The dynamical natures of the trace distance between the two states, $\cos\frac{\theta_1}{2}\ket{0}+\sin\frac{\theta_1}{2}\ket{1}$ and $\cos\frac{\theta_2}{2}\ket{0}+\sin\frac{\theta_2}{2}\ket{1}$, with $\theta_1+\theta_2=\frac{5\pi}{9}$, are plotted in the two panels. Panel (a) assumes coherent control over the lossy dynamics and we obtain IB at intermediate times, while panel (b) assumes the switch configuration for the same, where we obtain no IB. The vertical axes are dimensionless. The horizontal axes are also dimensionless, if we rescale time to time~$\times J/\hbar$, where $J$ is a multiplicative constant, of the $\gamma_i$ in the master equations, with the unit of energy.}
  \label{combined1}
\end{figure*}

This yields the trace distance between the pair of states given in Eq. (\ref{tr.dist.qs}) as
\begin{align*}
   \mathcal{D}[\rho_1^S(t),\rho_2^S(t)]=&\frac{1}{2}\|\vec{r}^S-\vec{s}^S\|\nonumber\\
   =&\frac{1}{2}\Big[N_{11}^2\Big((r_1-s_1)^2+(r_2-s_2)^2\Big)\nonumber\\
   &+N_{33}^2(r_3-s_3)^2\Big]^{\frac{1}{2}}
\end{align*}
Therefore,
\begin{align*}
  \frac{d}{dt}\mathcal{D}[\rho_1^S(t),\rho_2^S(t)]=-\frac{\xi^S(r_i,s_i,t)}{\mathcal{D}[\rho_1^S(t),\rho_2^S(t)]}
\end{align*}
where
\begin{align*}
  \xi^S(r_i,s_i,t)=&\frac{4e^{2t}(e^{4t}+6e^{2t}+1)}{(7e^{4t}+2e^{2t}-1)^2}\Big[N_{11}\Big((r_1-s_1)^2\\
  &+(r_2-s_2)^2\Big)
  +2N_{33}(r_3-s_3)^2\Big]
\end{align*}
\begin{center}
\begin{figure}[t!]
  \centering
    \includegraphics[width=.9\linewidth]{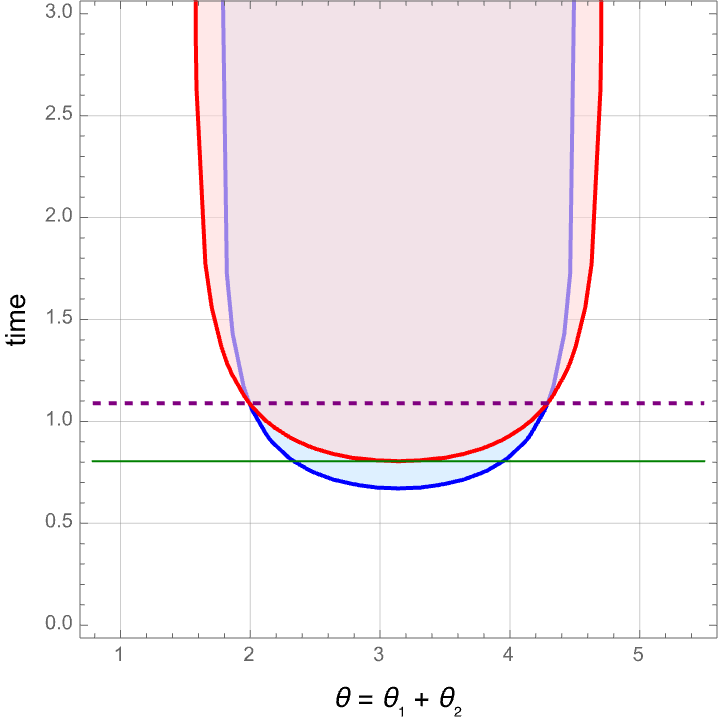}
\caption{Comparison of IB-activation regimes for initial states restricted to a common azimuthal plane. The vertical axis represents evolution time, while the horizontal axis denotes $\Theta=\theta_1+\theta_2$, the sum of Bloch-sphere polar angles for the two input states. Since, the IB admissible range is independent of $\Phi$ in the quantum-switch configuration (see Corollary \ref{cor_max_co_qs}), the \textcolor{blue}{blue} region marks the $(\Theta$, time$)$ pairs that activate IB for all azimuthal planes. The combined \textcolor{red}{red} region indicates the $(\Theta$, time$)$ pairs enabling IB in the coherently controlled path configuration on the $\Phi=0$ plane. Note that $\Phi=0$ yields the smallest IB-admitting region in this setup, and therefore serves as a lower-bound case for comparison. The solid green line corresponds to $t=\frac{1}{2}\ln 5$ while the dotted purple one represents $t_0=\frac{1}{2}\ln 8.85$. The plots show that although quantum switching allows a wider IB-admissible initial state space when $t\gtrsim\frac{1}{2}\ln 5$, the ordering reverses at later finite times and remains so asymptotically.}
  \label{combined2}
\end{figure}
\end{center}
In parallel with the analysis of interfering channels, the following proposition can be formulated for IB activation within the framework of the quantum switch.
\begin{proposition}\label{prop_max_co_sw}
    The pair of post-measurement states, given in Eq. (\ref{tr.dist.qs}), will exhibit IB at an instant $t$ if and only if
\begin{align*}
    (i)&~t>\frac{1}{2}\ln(1+2\sqrt{2})\\
    (ii)&~(r_3-s_3)^2>\frac{N_{11}[(r_1-s_1)^2
    +(r_2-s_2)^2]}{2|N_{33}|}
\end{align*}
\end{proposition}
\begin{proof}
    Similar to the proof of Proposition \ref{prop_max_co_ps}, here also we observe that to achieve our required condition, \textit{i.e.} $\frac{d}{dt}\mathcal{D}[\rho_1^S(t),\rho_2^S(t)]$ to be positive, \textit{i.e.} $\xi^S(r_i,s_i,t)<0$, we have
    \begin{align}\label{IB_qs}
        N_{11}\Big((r_1-s_1)^2+(r_2-s_2)^2\Big)+2N_{33}(r_3-s_3)^2<0
    \end{align}
    Hence, the only possibility here is $N_{33}$ to be negative, since all other terms are truly non-negative in $\xi^S(r_i,s_i,t)$. Hence, IB can be enabled \textit{only if}
    \begin{align*}
       e^{4t}-2e^{2t}-7>0 &\implies t>\frac{1}{2}\ln(1+2\sqrt{2})
    \end{align*}
    To prove the next condition, we again proceed in the similar way as in the proof of Proposition \ref{prop_max_co_ps}. Since, the desired time considers $t>\frac{1}{2}\ln(1+2\sqrt{2})$, we have to take $N_{33}=-|N_{33}|$ and thus the inequality (\ref{IB_qs}) turns out to be
    \begin{align*}
     &N_{11}\Big((r_1-s_1)^2+(r_2-s_2)^2\Big)-2|N_{33}|(r_3-s_3)^2<0\\
     \implies&(r_3-s_3)^2>\frac{N_{11}[(r_1-s_1)^2
    +(r_2-s_2)^2]}{2|N_{33}|}
    \end{align*}
    Hence, the \textit{if} condition is proved.
\end{proof}
Now, considering the Bloch vectors as given in Eq. (\ref{same_azi}), Proposition \ref{prop_max_co_sw} leads to the following corollary.
\begin{corollary}\label{cor_max_co_qs}
    All pair of post-measurement for which the Bloch vectors involves same $\Phi$, the condition of IB is as follows,
\begin{align*}
    \pi-\arccos\left(\frac{e^{4t}+6e^{2t}+17}{5e^{4t}-2e^{2t}-11}\right)<(\theta_1+\theta_2)\\
    <\pi+\arccos\left(\frac{e^{4t}+6e^{2t}+17}{5e^{4t}-2e^{2t}-11}\right)
\end{align*}
\end{corollary}
Note that quantum-switching triggers IB sooner in time. On the other hand, while determining which scheme permits a larger set of initial states to activate this phenomenon, the results established in Proposition \ref{prop_max_co_ps} and Proposition \ref{prop_max_co_sw} jointly imply the following :
\begin{theorem}\label{th_max_co}
    For every $\Phi$, there exists a critical point $t=t_0$ after which channel interference enables a broader set of initial states to exhibit IB compared to the quantum-switch configuration.
\end{theorem}
\begin{proof}
    Let us define two functions 
    \begin{align*}
       f(t,\Phi)&=\frac{e^{2t}-37-(e^{2t}-5)\cos 2\Phi}{11e^{2t}-23+(e^{2t}-5)\cos 2\Phi}\text{ and}\\
       g(t)&=\frac{e^{4t}+6e^{2t}+17}{5e^{4t}-2e^{2t}-11}.
    \end{align*}
    Note that the domain yielding the admissible values of $\theta_1+\theta_2$, spans an angular extent of $2\pi-2\arccos[f(t,\Phi)]=2\arccos[-f(t,\Phi)]$ in Corollary \ref{cor_max_co_ps}, whereas Corollary \ref{cor_max_co_qs} sets this extent to $2\arccos[g(t)]$. Since, $\arccos[x]$ is strictly decreasing on its domain, it suffices to compare the respective arguments in order to establish the above theorem.

We also define $h(t,\Phi)=-f(t,\Phi)-g(t)$. Evidently, at any time $t$, the maximum or minimum value of $h(t)$ depends only on $f(t,\Phi)$ as $g(t)$ depends only on $t$. Since $f(t,\Phi)$ possesses fractional affine form of $\cos 2\Phi$, its maximum or minimum value at a given $t$ is obtained at the end-points of $\cos 2\Phi$, that is at $\cos 2\Phi=\pm 1$. Furthermore, a simple anlytical derivation shows that $f(t,\Phi)$ is also a decreasing function in $\cos 2\Phi$. This readily implies that $-f(t,\Phi)$ is increasing in $\cos 2\Phi$. That means the maximum value of $-f(t,\Phi)$ is obtained for $\cos 2\Phi=1$. Therefore,
\begin{align*}
     h(t,0)&=-f(t,0)-g(t)\\
        &=-\frac{3e^{6t}-29e^{4t}+25e^{2t}-31}{(5e^{4t}-2e^{2t}-11)(3e^{2t}-7)}\\
        &=\frac{A(t)}{B(t)}
\end{align*}

Since, interfering of channels enables IB later in time, we will check the positivity of $h(t,0)$ in the range $t>\frac{1}{2}\ln 5$. Evidently, within this range, the denominator $B(t)$ is positive. Now for simplification, let us substitute $e^{2t}=x$ and we get 
\begin{align*}
    A(x)&=-3x^3+29x^2-25x+31\\
    \implies A^{\prime}(x)&=-9x^2+58x-25
    \end{align*}
Solving $A(x)=0$ yields two complex solutions and a single real solution located at $x\approx 8.85$. Evaluating the derivative at this point gives $A'(x)<0$ at $x\approx8.85$. Moreover, one can verify that $A'(x)$ remains negative throughout the interval $(8.85,\infty)$. Hence, $A(x)$ is strictly decreasing for all $x>8.85$. This in turn implies for all $t>\frac{1}{2}\ln 8.85=:t_0$,
\begin{align*}
    -f(t,0)&<g(t)\\
    \implies \arccos[-f(t)]&>\arccos[g(t)]
\end{align*}
 This completes the proof.
\end{proof}

In Theorem \ref{th_max_co} we analytically establish that two states lying on the same $\Phi$-plane admit a broader parameter range for IB activation under the coherent superposition–of–trajectories than under channel switching. This behavior is also demonstrated in Figs. \ref{combined1} and \ref{combined2}. Figure \ref{combined1} shows IB activation for a specific value of $\Theta=\theta_1+\theta_2$ in the interference configuration, whereas the same $\Theta$ fails to induce IB in the quantum-switch setup. Figure \ref{combined2} displays both the IB-activation instances and the corresponding ranges of admissible initial-state parameters under both the coherent-control configurations. It is important to emphasize that, although the quantum switch triggers IB earlier and initially admits a broader region of IB-compatible initial states, after a small time interval the superposition-of-trajectories framework surpasses it and supports IB over a larger subset of Bloch states. This inversion highlights a key point: an earlier onset of IB does not guarantee greater sustainability of initial-state variations throughout the evolution. In other words, temporal advantage and the sustainability of activation need not follow the same ordering in coherently-controlled open quantum dynamics.
\begin{figure*}[htbp]
  \centering
  \begin{subfigure}[b]{0.48\textwidth}
    \includegraphics[width=.9\linewidth]{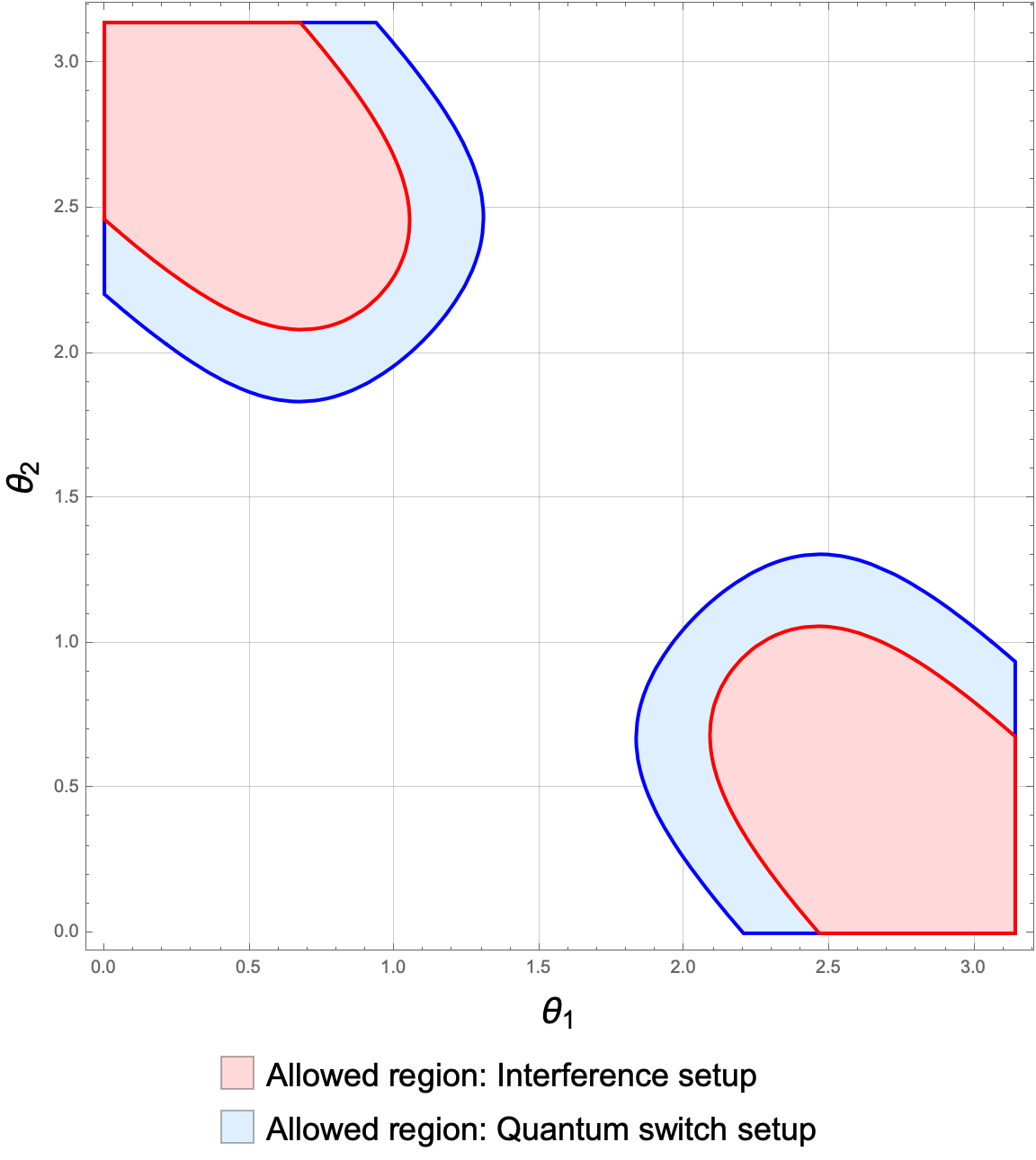}
    \caption{}
    \label{f1f2}
  \end{subfigure}
  \hfill
  \begin{subfigure}[b]{0.48\textwidth}
    \includegraphics[width=.9\linewidth]{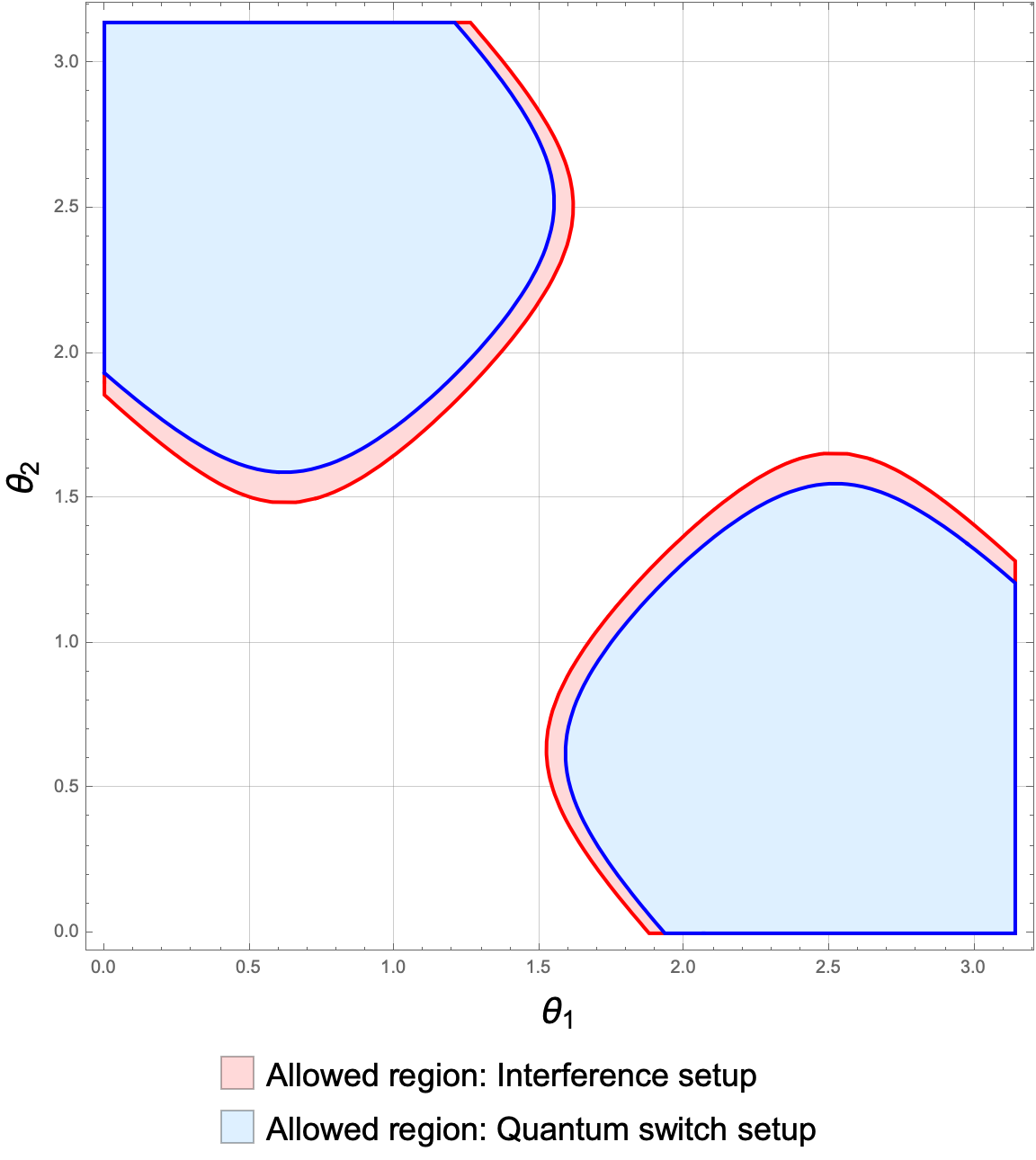}
    \caption{}
    \label{f2f1}
  \end{subfigure}
  \caption{Regions of the initial Bloch state space that allow IB activation are displayed at two finite time points, with $\Phi=\frac{\pi}{6}$ held fixed. (a) At $t=0.874719$, the quantum-switch setup supports IB for a wider subset of initial states than the interference setup. (b) At $t=1.20472$, this relation is exactly reversed, as the interference setup enables IB for a strictly larger initial-state region. This reversal emphasizes that earlier activation in time for some states does not ensure a persistent advantage across the full state space at later finite times.}
  \label{combined3}
\end{figure*}

In the preceding discussion, we restricted our attention to pairs of states lying on the same azimuthal plane. We will now consider a different instance in which the two states lie distinct azimuthal planes. Without loss of generality, we fix one state on the $\Phi=0$ plane and allow the other one to reside at an arbitrary azimuthal angle $\Phi$. The corresponding set of Bloch vectors are as follows :
\begin{align}\label{diff_azi}
        r_1&=\sin\theta_1 ~~&~~ s_1&=\sin\theta_2\cos\Phi\nonumber\\
        r_2&=0 ~~&~~ s_2&=\sin\theta_2\sin\Phi\nonumber\\
        r_3&=\cos\theta_1 ~~&~~ s_3&=\cos\theta_2
    \end{align}
    This different set from the previous one leads us to the following Observation.
 \begin{observation}\label{converse}
 For the pair of initial states specified in Eq. (\ref{diff_azi}), the quantum-switch configuration activates IB earlier and over a broader subset of states for times near $t\gtrsim\frac{1}{2}\ln 5$. However, this hierarchy does not persist as discussed in the  previous case. In a later, yet still finite, time regime  $t>>\frac{1}{2}\ln 5$, the interference configuration supports IB over a strictly larger fraction of the initial-state space at the same time instant.
\end{observation}
Note that we restrict our analysis to the time regime $t>\frac{1}{2}\ln 5$, as a meaningful comparison of the admissible state-space regions for the two coherently controlled configurations is possible only in this interval; for $t\leq\frac{1}{2}\ln 5$, the channel-interference setup does not permit IB activation. The set of initial Bloch states enabling IB activation under quantum-switch and the interference configurations is shown in Fig. \ref{combined3}, across two representative time scales. The figure depicts that the relative state-space support for IB activation can exhibit opposite hierarchies at different finite time intervals for both the controlled configurations of the channels.

\subsection{Robustness of coherent control}

Now, let us consider the control in a noisy state given as $\omega_p=p\ketbra{+}{+}+(1-p)\frac{\mathbb{I}}{2}$, where $p$ refers to the noise parameter and $0<p<1$. Evidently, $p=1$ boils down to the scenarios discussed above and $p=0$ makes the control maximally mixed which gives trivial solutions. However, the measurement on the control will be performed in the $\{\ketbra{\pm}{\pm}\}$ basis as we did before.

First we discuss the context of interfering $\phi_t$ with this noisy control. Here, the complex amplitudes $\{\alpha_i\}$ and $\{\beta_j\}$ take values as in the earlier case. Also, we select the post-measurement state of the target that corresponds to the $\ketbra{+}{+}$ outcome of the controller. Hence, after the channel action, the elements of the Bloch transfer matrix $M(p)$ of any output state $\rho^I_{p}(t)$ here is given as, 
\begin{align*}
M(p)_{ij}&=0~\forall i\neq j\text{ and }\\
    M(p)_{11}&=\frac{(2+p)e^{2t}+(2-p)}{e^{2t}(4+p)-p},\\
    M(p)_{22}&=\frac{(2-p)e^{2t}+(2+p)}{e^{2t}(4+p)-p},\\
    M(p)_{33}&=\frac{-pe^{2t}+(4+p)}{e^{2t}(4+p)-p}
\end{align*}
and the displacement vector here is $\vec{c}(p)=\vec{0}$.

Hence, the trace distance between the states $\{\rho^I_{p,i}(t)\}_{i=1}^2$ evolved from that given in Eq. (\ref{tr.dist.}) as
\begin{align*}
\mathcal{D}[\rho_{p,1}^I(t),\rho_{p,2}^I(t)]=&\frac{1}{2}\|\vec{r}^I(p)-\vec{s}^I(p)\|\nonumber\\
   =&\frac{1}{2}\Big[M(p)_{11}^2(r_1-s_1)^2+M(p)_{22}^2(r_2-s_2)^2\nonumber\\
   &+M(p)_{33}^2(r_3-s_3)^2\Big]^{\frac{1}{2}}
\end{align*}
Therefore,
\begin{align*}
  \frac{d}{dt}\mathcal{D}[\rho_{p,1}^I(t),\rho_{p,2}^I(t)]=-\frac{\xi_p^I(r_i,s_i,t)}{\mathcal{D}[\rho_{p,1}^I(t),\rho_{p,2}^I(t)]}
\end{align*}
where
\begin{align*}
  \xi^I_p(r_i,s_i,t)=&\frac{4e^{2t}}{[e^{2t}(4+p)-p]^2}\Big[M(p)_{11}(r_1-s_1)^2\\
    &+(1+p)M(p)_{22}(r_2-s_2)^2\\
    &+(2+p)M(p)_{33}(r_3-s_3)^2\Big]
\end{align*}

\begin{center}
\begin{figure*}[!htb]
  \centering
  \begin{subfigure}[b]{0.48\textwidth}
    \includegraphics[width=.9\linewidth]{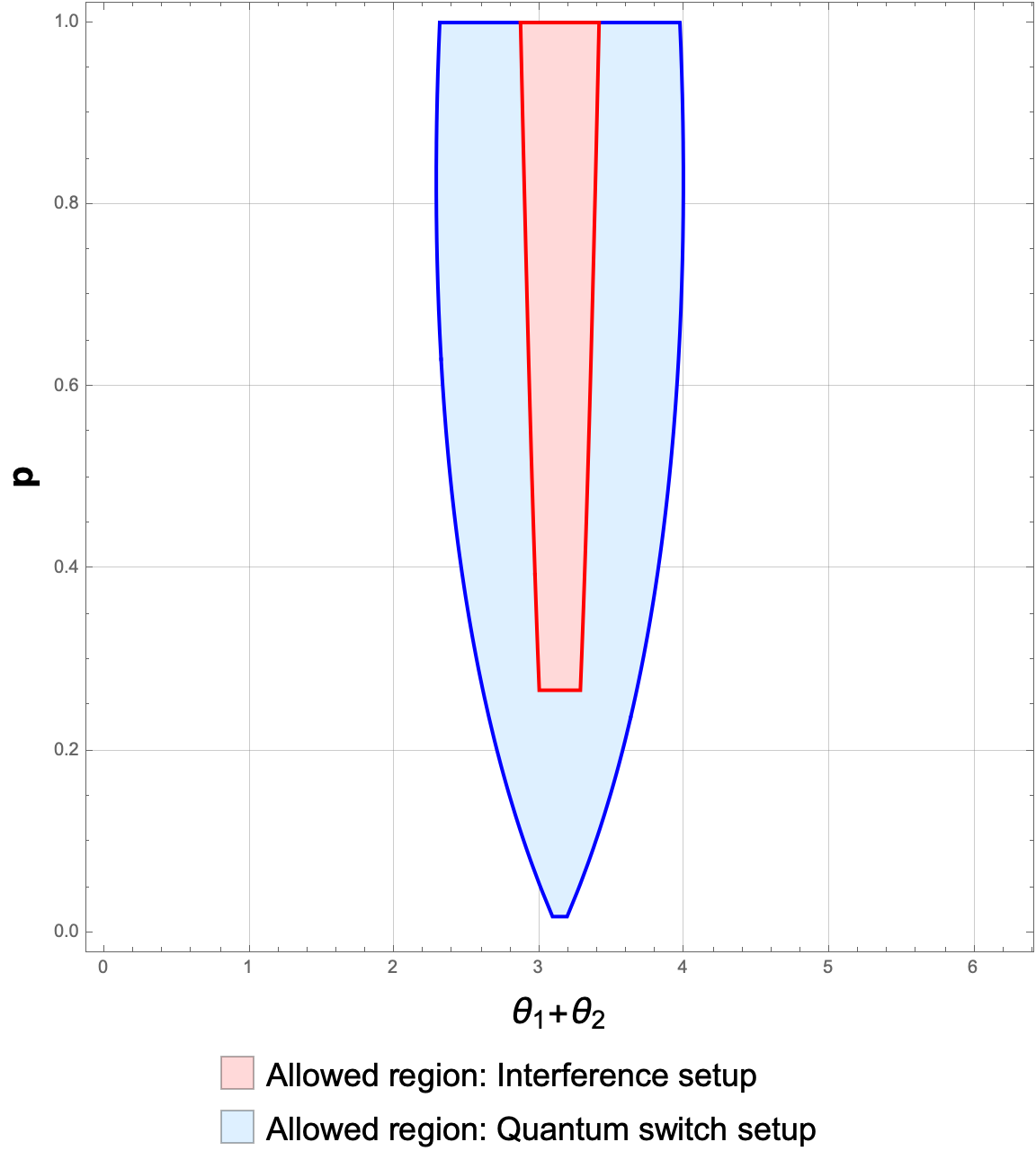}
    \caption{}
    \label{pf1f2}
  \end{subfigure}
  \hfill
  \begin{subfigure}[b]{0.48\textwidth}
    \includegraphics[width=.9\linewidth]{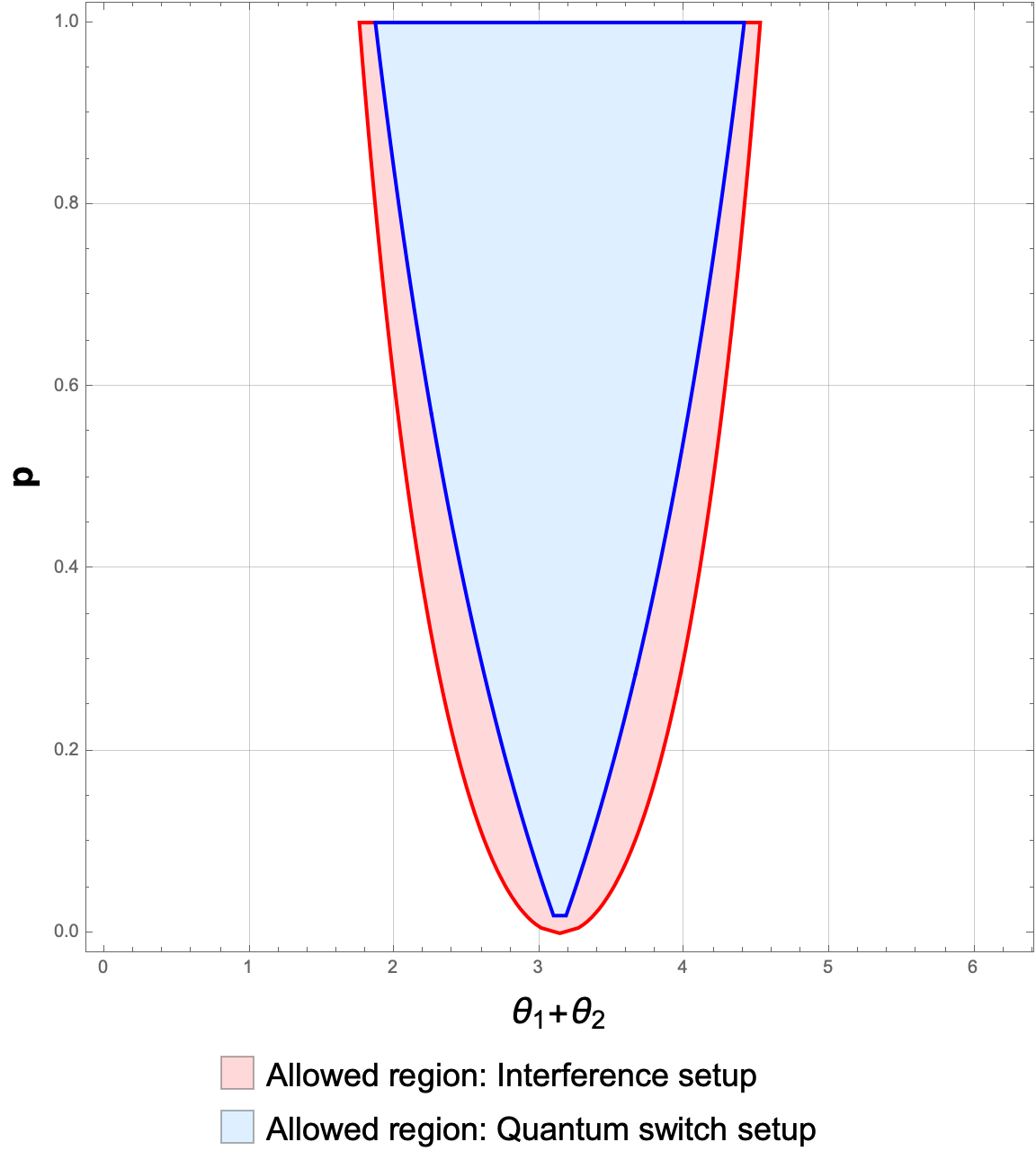}
    \caption{}
    \label{pf2f1}
  \end{subfigure}
\caption{Comparison of IB-supporting regions at different finite time instances, with both initial states restricted to the $xz$-plane of the Bloch sphere under noisy control where $p$ denotes the noise parameter. (a) At $t=\frac{1}{2}\ln\left(1+\frac{4}{p}\right)+0.01$, the quantum switch enables IB over a larger initial-state region and with enhanced robustness in control than is possible in the interference-based setup. (b) At $t=\frac{1}{2}\ln\left(1+\frac{4}{p}\right)+0.6$, the previous  ordering is reversed, as the interference setup admits IB for a strictly larger set of initial states and under more robust control conditions.
}
  \label{combined4}
\end{figure*}
\end{center}

We are now in a position to state the next Proposition under this current setup.
\begin{proposition}\label{prop_rc_ps}
    The pair of post-measurement states $\{\rho^I_{p,i}(t)\}_{i=1}^2$, under the interference of the channel $\phi_t$ with noisy control, will exhibit IB at an instant $t$ \textit{if and only if}
\begin{align*}
    (i)&~t>\frac{1}{2}\ln\left(1+\frac{4}{p}\right)\\
    (ii)&~(r_3-s_3)^2>\\
    &\qquad\frac{M(p)_{11}(r_1-s_1)^2
    +(1+p)M(p)_{22}(r_2-s_2)^2}{(2+p)|M(p)_{33}|}
\end{align*}
\end{proposition}
\begin{proof}
    As before, the condition to activate IB demands $\xi^I_p(r_i,s_i,t)<0$ and thus $M(p)_{33}<0$. Hence,
    \begin{align*}
        -pe^{2t}+(4+p)<0 &\implies t>\frac{1}{2}\ln\left(1+\frac{4}{p}\right)
    \end{align*}
    Also, $\xi^I_p(r_i,s_i,t)<0$ leads to the following condition
    \begin{align*}
       M(p)_{11}(r_1-s_1)^2&+(1+p)M(p)_{22}(r_2-s_2)^2\\
       &+(2+p)M(p)_{33}(r_3-s_3)^2 <0
    \end{align*}
    Following exactly same arguments as before, we arrive
    \begin{align*}
       (r_3-s_3)^2&>\\
    &\frac{M(p)_{11}(r_1-s_1)^2
    +(1+p)M(p)_{22}(r_2-s_2)^2}{(2+p)|M(p)_{33}|} 
    \end{align*}
\end{proof}

Now consider the set of initial Bloch vectors as given in Eq. (\ref{same_azi}) with $\Phi=0$. These initial states along with Proposition \ref{prop_rc_ps} lead to the following Corollary.
\begin{corollary}\label{cor_rc_ps}
    The condition for activating IB with a noisy control is satisfied if the state parameter satisfies
    \begin{align*}
  \pi-\arccos\left(\frac{ae^{2t}+b}{ce^{2t}+d}\right)&<(\theta_1+\theta_2)\\
    &<\pi+\arccos\left(\frac{ae^{2t}+b}{ce^{2t}+d}\right)\\
    \text{where,}\qquad\qquad\qquad&\\
    a=2-p-p^2,&~b=10+5p+p^2\\
    c=(2+p)(1+p),&~d=-(6+p)(1+p)  
\end{align*}
\end{corollary}

Next, we consider the another coherently controlled configuration, aka quantum-switch, with the control $\omega_p$. Keeping all the setup as before, we obtain the elements of the Bloch transfer matrix $N(p)$ of any evolved state $\rho^S_p(t)$ as 
\begin{align*}
        &N(p)_{ij}=0~\forall i\neq j;\\
        &N(p)_{11}=N(p)_{22}=\frac{(1+2p)(e^{4t}+1)+2e^{2t}}{(4+3p)e^{4t}+2pe^{2t}-p},\\
        &N(p)_{33}=-\frac{pe^{4t}-2pe^{2t}-(3p+4)}{(4+3p)e^{4t}+2pe^{2t}-p}.
\end{align*}
and the displacement vector $\vec{d}(p)=0$. Note that in this case as well, rotational symmetry about the $z$-axis is preserved, as in the maximal coherent control configuration.

Therefore, we obtain the change in the trace distance between the states $\{\rho^S_{p,i}(t)\}_{i=1}^2$ evolved from that given in Eq. (\ref{tr.dist.}) as
\begin{align*}
  \frac{d}{dt}\mathcal{D}[\rho_{p,1}^S(t),\rho_{p,2}^S(t)]=-\frac{\xi^S_p(r_i,s_i,t)}{\mathcal{D}[\rho_{p,1}^S(t),\rho_{p,2}^S(t)]}
\end{align*}
where the trace distance is given by, 
\begin{align*}
\mathcal{D}[\rho_{p,1}^S(t),\rho_{p,2}^S(t)]=&\frac{1}{2}\|\vec{r}^S(p)-\vec{s}^S(p)\|\\
   =&\frac{1}{2}\Big[N(p)_{11}^2\Big((r_1-s_1)^2+(r_2-s_2)^2\Big)\\
   &+N(p)_{33}^2(r_3-s_3)^2\Big]^{\frac{1}{2}}\\
   \text{and}\qquad\qquad\qquad&\\
  \xi^S_p(r_i,s_i,t)=&\frac{2e^{2t}(1+p)}{[(4+3p)e^{4t}+2pe^{2t}-p]^2}\times\qquad\\
  \Big[f_1(p,t)
  &N(p)_{11}\Big((r_1-s_1)^2+(r_2-s_2)^2\Big)\\
  &+2f_2(p,t)N(p)_{33}(r_3-s_3)^2\Big]\\
  \text{with}\qquad\qquad\qquad&\\
  f_1(p,t)&=(2-p)e^{4t}+(2+4p)e^{2t}+p\\
    f_2(p,t)&=pe^{4t}+2(2+p)e^{2t}+p
\end{align*}
This leads to the following result.
\begin{proposition}\label{prop_rc_qs}
    The pair of post-measurement states $\{\rho^S_{p,i}(t)\}_{i=1}^2$, under the configuration of quantum-switch with noisy control, will exhibit IB at an instant $t$ \textit{if and only if}
\begin{align*}
    (i)&~t>\frac{1}{2}\ln\left(1+2\sqrt{1+\frac{1}{p}}\right)\\
    (ii)&~(r_3-s_3)^2>\\
    &\qquad\frac{f_1(p,t)N(p)_{11}\Big((r_1-s_1)^2
    +(r_2-s_2)^2\Big)}{2f_2(p,t)|N(p)_{33}|}
\end{align*}
\end{proposition}
To prove Proposition \ref{prop_rc_qs}, we need $\xi^S_p(r_i,s_i,t)<0$.\\
This readily implies 
\begin{align*}
  N(p)_{33}<0~\text{ and hence, }~t>\frac{1}{2}\ln\left(1+2\sqrt{1+\frac{1}{p}}\right)
\end{align*}
 Following the same outline of the earlier proofs, the other condition  can also be established.

Now, if we consider two initial states lying on the $xz$-plane of the Bloch sphere, Proposition \ref{prop_rc_qs} yields the following Corollary, providing bounds on the corresponding state parameters.

\begin{corollary}\label{cor_rc_qs}
     IB can be activated with the initial states on the $xz$-plane under the framework of quantum-switch if the state parameters satisfy the following inequality 
\begin{align*}
    \pi-\arccos&\left(\frac{\sum_{j=0}^4 A_je^{2jt}}{A_5}\right)<(\theta_1+\theta_2)\\&<\pi+\arccos\left(\frac{\sum_{j=0}^4 A_je^{2jt}}{A_5}\right)
    \intertext{where,}\\
    A_0&=p(9+8p)\\
A_1&=2(17+25p+12p^2)\\
A_2&=6(1+6p+2p^2)\\
A_3&=2(3-p+4p^2)\\
A_4&=2+3p-4p^2\\
A_5&=e^{8t}(2+3p)+2e^{6t}(1+p)(3+4p)\\
&+6e^{4t}(1-2p-2p^2)-2e^{2t}(15+15p+4p^2)\\
&-p(7+4p)
\end{align*}
\end{corollary}

Figure \ref{combined4} illustrates an inversion in the ordering of both the admissible initial-state space and the degree of robustness in the control input across two finite time instances required for IB to occur. We restrict our analysis to the time regime $t>\frac{1}{2}\ln\left(1+\frac{4}{p}\right)$ to enable a meaningful comparison, as discussed previously. The reversed ordering observed in panels \ref{pf1f2} and \ref{pf2f1} closely mirrors the behavior described in Observation \ref{converse}. 


\section{Conclusion}

In summary, we have shown that interfering two lossy non-Markovian dynamics can lead to 
instances of IB. While a similar situation may arise by placing these trajectories in an indefinite causal order, aka quantum-switch, the controlled configuration over the choice of trajectories offers a higher efficacy. Precisely, we have shown that  interference of the lossy trajectories enables IB for a substantially wider class of quantum states. Moreover, 
the amount of white noise tolerated by the controlling device to depict the IB is higher for the superposed quantum operation trajectories configuration than that of the quantum-switch configuration.

Notably, while the quantum-switch can outperform the coherently controlled choice configuration in certain quantum communication tasks \cite{chiribella2021}, there also exist contexts where both control schemes yield equivalent advantages \cite{abbott2020}.
Our result however forms an instance 
where coherent control over the choice of trajectories outperforms the quantum-switch setup, both in the ideal and in the noisy cases. 
However, it remains an interesting direction for future investigation that what key features of lossy dynamics allow coherent control over trajectory selection to surpass the performance of the quantum-switch. Another interesting open direction that remains unanswered is whether such an efficient reversal of information, via interference, actually persists for any general class of lossy dynamics? Or, do we need to demand at least an element of non-Markovianity beforehand in the dynamics itself?

\bibliography{bibliography}

@article{abbott2020,
  doi = {10.22331/q-2020-09-24-333},
  url = {https://doi.org/10.22331/q-2020-09-24-333},
  title = {Communication through coherent control of quantum channels},
  author = {Abbott, Alastair A. and Wechs, Julian and Horsman, Dominic and Mhalla, Mehdi and Branciard, Cyril},
  journal = {{Quantum}},
  issn = {2521-327X},
  publisher = {{Verein zur F{\"{o}}rderung des Open Access Publizierens in den Quantenwissenschaften}},
  volume = {4},
  pages = {333},
  month = sep,
  year = {2020}
}

@article{aberg2004,
  title = {Operations and single-particle interferometry},
  author = {\AA{}berg, Johan},
  journal = {Phys. Rev. A},
  volume = {70},
  issue = {1},
  pages = {012103},
  numpages = {10},
  year = {2004},
  month = {Jul},
  publisher = {American Physical Society},
  doi = {10.1103/PhysRevA.70.012103},
  url = {https://link.aps.org/doi/10.1103/PhysRevA.70.012103}
}

@article{aharonov1990,
  title = {Superpositions of time evolutions of a quantum system and a quantum time-translation machine},
  author = {Aharonov, Yakir and Anandan, Jeeva and Popescu, Sandu and Vaidman, Lev},
  journal = {Phys. Rev. Lett.},
  volume = {64},
  issue = {25},
  pages = {2965--2968},
  numpages = {0},
  year = {1990},
  month = {Jun},
  publisher = {American Physical Society},
  doi = {10.1103/PhysRevLett.64.2965},
  url = {https://link.aps.org/doi/10.1103/PhysRevLett.64.2965}
}

@artucle{ambainis2008,
      title={Quantum Random Access Codes with Shared Randomness}, 
      author={Andris Ambainis and Debbie Leung and Laura Mancinska and Maris Ozols},
      year={2009},
      eprint={0810.2937},
      archivePrefix={arXiv},
      primaryClass={quant-ph},
      url={https://arxiv.org/abs/0810.2937}
}

@article{andersson2007,
author = {Erika Andersson and James D. Cresser and Michael J. W. Hall},
title = {Finding the Kraus decomposition from a master equation and vice versa},
journal = {Journal of Modern Optics},
volume = {54},
number = {12},
pages = {1695--1716},
year = {2007},
publisher = {Taylor \& Francis},
doi = {10.1080/09500340701352581},
URL = {https://doi.org/10.1080/09500340701352581},
eprint = {https://doi.org/10.1080/09500340701352581}
}

@misc{benabdallah2025,
      title={Non-Markovian Protection and Thermal Fragility of Quantum Resources in a Spin-1/2 Ising-Heisenberg Diamond Chain}, 
      author={Fadwa Benabdallah and M. Y. Abd-Rabbou and Mohammed Daoud},
      year={2025},
      eprint={2506.12734},
      archivePrefix={arXiv},
      primaryClass={quant-ph},
      url={https://arxiv.org/abs/2506.12734}, 
}

@article{benatti2024,
  title = {Quantum versus classical $P$-divisibility},
  author = {Benatti, Fabio and Chru\ifmmode \acute{s}\else \'{s}\fi{}ci\ifmmode \acute{n}\else \'{n}\fi{}ski, Dariusz and Nichele, Giovanni},
  journal = {Phys. Rev. A},
  volume = {110},
  issue = {5},
  pages = {052212},
  numpages = {13},
  year = {2024},
  month = {Nov},
  publisher = {American Physical Society},
  doi = {10.1103/PhysRevA.110.052212},
  url = {https://link.aps.org/doi/10.1103/PhysRevA.110.052212}
}

@article{bennett1984,
title = {Quantum cryptography: Public key distribution and coin tossing},
journal = {Theoretical Computer Science},
volume = {560},
pages = {7-11},
year = {2014},
note = {Theoretical Aspects of Quantum Cryptography – celebrating 30 years of BB84},
issn = {0304-3975},
doi = {https://doi.org/10.1016/j.tcs.2014.05.025},
url = {https://www.sciencedirect.com/science/article/pii/S0304397514004241},
author = {Charles H. Bennett and Gilles Brassard}
}

@article{bhattacharya2021,
  title = {Random-Receiver Quantum Communication},
  author = {Bhattacharya, Some Sankar and Maity, Ananda G. and Guha, Tamal and Chiribella, Giulio and Banik, Manik},
  journal = {PRX Quantum},
  volume = {2},
  issue = {2},
  pages = {020350},
  numpages = {13},
  year = {2021},
  month = {Jun},
  publisher = {American Physical Society},
  doi = {10.1103/PRXQuantum.2.020350},
  url = {https://link.aps.org/doi/10.1103/PRXQuantum.2.020350}
}

@article{blondeau2021,
  title = {Noisy quantum metrology with the assistance of indefinite causal order},
  author = {Chapeau-Blondeau, Fran\ifmmode \mbox{\c{c}}\else \c{c}\fi{}ois},
  journal = {Phys. Rev. A},
  volume = {103},
  issue = {3},
  pages = {032615},
  numpages = {18},
  year = {2021},
  month = {Mar},
  publisher = {American Physical Society},
  doi = {10.1103/PhysRevA.103.032615},
  url = {https://link.aps.org/doi/10.1103/PhysRevA.103.032615}
}

@book{breuer2002,
  title={The theory of open quantum systems},
  author={Breuer, Heinz-Peter and Petruccione, Francesco},
  year={2002},
  publisher={OUP Oxford}
}

@article{breuer2009,
  title = {Measure for the Degree of Non-Markovian Behavior of Quantum Processes in Open Systems},
  author = {Breuer, Heinz-Peter and Laine, Elsi-Mari and Piilo, Jyrki},
  journal = {Phys. Rev. Lett.},
  volume = {103},
  issue = {21},
  pages = {210401},
  numpages = {4},
  year = {2009},
  month = {Nov},
  publisher = {American Physical Society},
  doi = {10.1103/PhysRevLett.103.210401},
  url = {https://link.aps.org/doi/10.1103/PhysRevLett.103.210401}
}

@article{bylicka2014,
  title={Non-Markovianity and reservoir memory of quantum channels: a quantum information theory perspective},
  author={Bylicka, Bogna and Chru{\'s}ci{\'n}ski, D and Maniscalco, Sci},
  journal={Scientific reports},
  volume={4},
  number={1},
  pages={5720},
  year={2014},
  doi={https://doi.org/10.1038/srep05720},
  publisher={Nature Publishing Group UK London}
}

@article{chiribella2013,
  title = {Quantum computations without definite causal structure},
  author = {Chiribella, Giulio and D'Ariano, Giacomo Mauro and Perinotti, Paolo and Valiron, Benoit},
  journal = {Phys. Rev. A},
  volume = {88},
  issue = {2},
  pages = {022318},
  numpages = {15},
  year = {2013},
  month = {Aug},
  publisher = {American Physical Society},
  doi = {10.1103/PhysRevA.88.022318},
  url = {https://link.aps.org/doi/10.1103/PhysRevA.88.022318}
}

@article{chiribella2019,
   title={Quantum Shannon theory with superpositions of trajectories},
   volume={475},
   ISSN={1471-2946},
   url={http://dx.doi.org/10.1098/rspa.2018.0903},
   DOI={10.1098/rspa.2018.0903},
   number={2225},
   journal={Proceedings of the Royal Society A: Mathematical, Physical and Engineering Sciences},
   publisher={The Royal Society},
   author={Chiribella, Giulio and Kristjánsson, Hlér},
   year={2019},
   month=may, pages={20180903} 
}

@article{chiribella2021,
doi = {10.1088/1367-2630/abe7a0},
url = {https://dx.doi.org/10.1088/1367-2630/abe7a0},
year = {2021},
month = {mar},
publisher = {IOP Publishing},
volume = {23},
number = {3},
pages = {033039},
author = {Chiribella, Giulio and Banik, Manik and Bhattacharya, Some Sankar and Guha, Tamal and Alimuddin, Mir and Roy, Arup and Saha, Sutapa and Agrawal, Sristy and Kar, Guruprasad},
title = {Indefinite causal order enables perfect quantum communication with zero capacity channels},
journal = {New Journal of Physics}
}

@article{chiribella2021a,
  title = {Quantum and Classical Data Transmission through Completely Depolarizing Channels in a Superposition of Cyclic Orders},
  author = {Chiribella, Giulio and Wilson, Matt and Chau, H. F.},
  journal = {Phys. Rev. Lett.},
  volume = {127},
  issue = {19},
  pages = {190502},
  numpages = {6},
  year = {2021},
  month = {Nov},
  publisher = {American Physical Society},
  doi = {10.1103/PhysRevLett.127.190502},
  url = {https://link.aps.org/doi/10.1103/PhysRevLett.127.190502}
}

@article{chruscinski2014,
  title = {Degree of Non-Markovianity of Quantum Evolution},
  author = {Chru\ifmmode \acute{s}\else \'{s}\fi{}ci\ifmmode \acute{n}\else \'{n}\fi{}ski, Dariusz and Maniscalco, Sabrina},
  journal = {Phys. Rev. Lett.},
  volume = {112},
  issue = {12},
  pages = {120404},
  numpages = {5},
  year = {2014},
  month = {Mar},
  publisher = {American Physical Society},
  doi = {10.1103/PhysRevLett.112.120404},
  url = {https://link.aps.org/doi/10.1103/PhysRevLett.112.120404}
}

@article{dąbrowska2021,
doi = {10.1088/1367-2630/ac3c60},
url = {https://dx.doi.org/10.1088/1367-2630/ac3c60},
year = {2021},
month = {dec},
publisher = {IOP Publishing},
volume = {23},
number = {12},
pages = {123019},
author = {Dąbrowska, Anita and Chruściński, Dariusz and Chakraborty, Sagnik and Sarbicki, Gniewomir},
title = {Eternally non-Markovian dynamics of a qubit interacting with a single-photon wavepacket},
journal = {New Journal of Physics}
}

@article{deutsch1992,
 ISSN = {09628444},
 URL = {https://doi.org/10.1098/rspa.1992.0167},
 author = {David Deutsch and Richard Jozsa},
 journal = {Proceedings: Mathematical and Physical Sciences},
 number = {1907},
 pages = {553--558},
 publisher = {Royal Society},
 title = {Rapid Solution of Problems by Quantum Computation},
 urldate = {2025-07-26},
 volume = {439},
 year = {1992}
}

@article{ebler2018,
  title = {Enhanced Communication with the Assistance of Indefinite Causal Order},
  author = {Ebler, Daniel and Salek, Sina and Chiribella, Giulio},
  journal = {Phys. Rev. Lett.},
  volume = {120},
  issue = {12},
  pages = {120502},
  numpages = {5},
  year = {2018},
  month = {Mar},
  publisher = {American Physical Society},
  doi = {10.1103/PhysRevLett.120.120502},
  url = {https://link.aps.org/doi/10.1103/PhysRevLett.120.120502}
}

@article{farkas2025,
  doi = {10.22331/q-2025-02-25-1643},
  url = {https://doi.org/10.22331/q-2025-02-25-1643},
  title = {Simple and general bounds on quantum random access codes},
  author = {Farkas, M{\'{a}}t{\'{e}} and Miklin, Nikolai and Tavakoli, Armin},
  journal = {{Quantum}},
  issn = {2521-327X},
  publisher = {{Verein zur F{\"{o}}rderung des Open Access Publizierens in den Quantenwissenschaften}},
  volume = {9},
  pages = {1643},
  month = feb,
  year = {2025}
}

@article{felce2020,
  title = {Quantum Refrigeration with Indefinite Causal Order},
  author = {Felce, David and Vedral, Vlatko},
  journal = {Phys. Rev. Lett.},
  volume = {125},
  issue = {7},
  pages = {070603},
  numpages = {6},
  year = {2020},
  month = {Aug},
  publisher = {American Physical Society},
  doi = {10.1103/PhysRevLett.125.070603},
  url = {https://link.aps.org/doi/10.1103/PhysRevLett.125.070603}
}

@Article{giorgi2020,
AUTHOR = {Giorgi, Gian Luca and Lorenzo, Salvatore and Longhi, Stefano},
TITLE = {Topological Protection and Control of Quantum Markovianity},
JOURNAL = {Photonics},
VOLUME = {7},
YEAR = {2020},
NUMBER = {1},
ARTICLE-NUMBER = {18},
URL = {https://www.mdpi.com/2304-6732/7/1/18},
ISSN = {2304-6732},
doi = {10.3390/photonics7010018}
}

@article{gisin2005,
  title = {Error filtration and entanglement purification for quantum communication},
  author = {Gisin, N. and Linden, N. and Massar, S. and Popescu, S.},
  journal = {Phys. Rev. A},
  volume = {72},
  issue = {1},
  pages = {012338},
  numpages = {17},
  year = {2005},
  month = {Jul},
  publisher = {American Physical Society},
  doi = {10.1103/PhysRevA.72.012338},
  url = {https://link.aps.org/doi/10.1103/PhysRevA.72.012338}
}

@article{gorini1978,
title = {Properties of quantum Markovian master equations},
journal = {Reports on Mathematical Physics},
volume = {13},
number = {2},
pages = {149-173},
year = {1978},
issn = {0034-4877},
doi = {https://doi.org/10.1016/0034-4877(78)90050-2},
url = {https://www.sciencedirect.com/science/article/pii/0034487778900502},
author = {Vittorio Gorini and Alberto Frigerio and Maurizio Verri and Andrzej Kossakowski and E.C.G. Sudarshan}
}

@article{goswami2018,
  title = {Indefinite Causal Order in a Quantum Switch},
  author = {Goswami, K. and Giarmatzi, C. and Kewming, M. and Costa, F. and Branciard, C. and Romero, J. and White, A. G.},
  journal = {Phys. Rev. Lett.},
  volume = {121},
  issue = {9},
  pages = {090503},
  numpages = {5},
  year = {2018},
  month = {Aug},
  publisher = {American Physical Society},
  doi = {10.1103/PhysRevLett.121.090503},
  url = {https://link.aps.org/doi/10.1103/PhysRevLett.121.090503}
}

@article{guha2020,
  title = {Thermodynamic advancement in the causally inseparable occurrence of thermal maps},
  author = {Guha, Tamal and Alimuddin, Mir and Parashar, Preeti},
  journal = {Phys. Rev. A},
  volume = {102},
  issue = {3},
  pages = {032215},
  numpages = {6},
  year = {2020},
  month = {Sep},
  publisher = {American Physical Society},
  doi = {10.1103/PhysRevA.102.032215},
  url = {https://link.aps.org/doi/10.1103/PhysRevA.102.032215}
}

@article{guha2023,
  title = {Quantum networks boosted by entanglement with a control system},
  author = {Guha, Tamal and Roy, Saptarshi and Chiribella, Giulio},
  journal = {Phys. Rev. Res.},
  volume = {5},
  issue = {3},
  pages = {033214},
  numpages = {16},
  year = {2023},
  month = {Sep},
  publisher = {American Physical Society},
  doi = {10.1103/PhysRevResearch.5.033214},
  url = {https://link.aps.org/doi/10.1103/PhysRevResearch.5.033214}
}

@article{guo2020,
  title = {Experimental Transmission of Quantum Information Using a Superposition of Causal Orders},
  author = {Guo, Yu and Hu, Xiao-Min and Hou, Zhi-Bo and Cao, Huan and Cui, Jin-Ming and Liu, Bi-Heng and Huang, Yun-Feng and Li, Chuan-Feng and Guo, Guang-Can and Chiribella, Giulio},
  journal = {Phys. Rev. Lett.},
  volume = {124},
  issue = {3},
  pages = {030502},
  numpages = {6},
  year = {2020},
  month = {Jan},
  publisher = {American Physical Society},
  doi = {10.1103/PhysRevLett.124.030502},
  url = {https://link.aps.org/doi/10.1103/PhysRevLett.124.030502}
}

@misc{hadipour2025,
      title={Quantum Speed Limits in Qubit Dynamics Driven by Bistable Random Telegraph Noise: From Markovian to Non-Markovian Regimes}, 
      author={Maryam Hadipour},
      year={2025},
      eprint={2504.15808},
      archivePrefix={arXiv},
      primaryClass={quant-ph},
      url={https://arxiv.org/abs/2504.15808}, 
}

@article{hall2008,
doi = {10.1088/1751-8113/41/20/205302},
url = {https://dx.doi.org/10.1088/1751-8113/41/20/205302},
year = {2008},
month = {apr},
publisher = {},
volume = {41},
number = {20},
pages = {205302},
author = {Hall, Michael J W},
title = {Complete positivity for time-dependent qubit master equations},
journal = {Journal of Physics A: Mathematical and Theoretical}
}

@article{hall2014,
  title = {Canonical form of master equations and characterization of non-Markovianity},
  author = {Hall, Michael J. W. and Cresser, James D. and Li, Li and Andersson, Erika},
  journal = {Phys. Rev. A},
  volume = {89},
  issue = {4},
  pages = {042120},
  numpages = {11},
  year = {2014},
  month = {Apr},
  publisher = {American Physical Society},
  doi = {10.1103/PhysRevA.89.042120},
  url = {https://link.aps.org/doi/10.1103/PhysRevA.89.042120}
}

@article{hardy2007,
doi = {10.1088/1751-8113/40/12/S12},
url = {https://dx.doi.org/10.1088/1751-8113/40/12/S12},
year = {2007},
month = {mar},
publisher = {},
volume = {40},
number = {12},
pages = {3081},
author = {Hardy, Lucien},
title = {Towards quantum gravity: a framework for probabilistic theories with non-fixed causal structure},
journal = {Journal of Physics A: Mathematical and Theoretical}
}

@book{heinosaari2011,
  title={The mathematical language of quantum theory: from uncertainty to entanglement},
  author={Heinosaari, Teiko and Ziman, M{\'a}rio},
  year={2011},
  publisher={Cambridge University Press}
}

@article{jahromi2019,
   title={Multiparameter estimation, lower bound on quantum Fisher information, and non-Markovianity witnesses of noisy two-qubit systems},
   volume={18},
   ISSN={1573-1332},
   url={http://dx.doi.org/10.1007/s11128-019-2446-8},
   DOI={10.1007/s11128-019-2446-8},
   number={11},
   journal={Quantum Information Processing},
   publisher={Springer Science and Business Media LLC},
   author={Jahromi, H. Rangani and Amini, M. and Ghanaatian, M.},
   year={2019},
   month=sep }

@article{khurana2019,
  title = {Experimental emulation of quantum non-Markovian dynamics and coherence protection in the presence of information backflow},
  author = {Khurana, Deepak and Agarwalla, Bijay Kumar and Mahesh, T. S.},
  journal = {Phys. Rev. A},
  volume = {99},
  issue = {2},
  pages = {022107},
  numpages = {7},
  year = {2019},
  month = {Feb},
  publisher = {American Physical Society},
  doi = {10.1103/PhysRevA.99.022107},
  url = {https://link.aps.org/doi/10.1103/PhysRevA.99.022107}
}

@article{laine2014,
  title={Nonlocal memory effects allow perfect teleportation with mixed states},
  author={Laine, Elsi-Mari and Breuer, Heinz-Peter and Piilo, Jyrki},
  journal={Scientific reports},
  volume={4},
  number={1},
  pages={4620},
  year={2014},
  doi={https://doi.org/10.1038/srep04620},
  publisher={Nature Publishing Group UK London}
}

@article{lamoureux2005,
  title = {Experimental Error Filtration for Quantum Communication Over Highly Noisy Channels},
  author = {Lamoureux, L.-P. and Brainis, E. and Cerf, N. J. and Emplit, Ph. and Haelterman, M. and Massar, S.},
  journal = {Phys. Rev. Lett.},
  volume = {94},
  issue = {23},
  pages = {230501},
  numpages = {4},
  year = {2005},
  month = {Jun},
  publisher = {American Physical Society},
  doi = {10.1103/PhysRevLett.94.230501},
  url = {https://link.aps.org/doi/10.1103/PhysRevLett.94.230501}
}

@article{lai2024,
  title = {Quick charging of a quantum battery with superposed trajectories},
  author = {Lai, Po-Rong and Lin, Jhen-Dong and Huang, Yi-Te and Jan, Hsien-Chao and Chen, Yueh-Nan},
  journal = {Phys. Rev. Res.},
  volume = {6},
  issue = {2},
  pages = {023136},
  numpages = {15},
  year = {2024},
  month = {May},
  publisher = {American Physical Society},
  doi = {10.1103/PhysRevResearch.6.023136},
  url = {https://link.aps.org/doi/10.1103/PhysRevResearch.6.023136}
}

@article{lindblad1976,
  title={On the generators of quantum dynamical semigroups},
  author={Lindblad, Goran},
  journal={Communications in mathematical physics},
  volume={48},
  number={2},
  pages={119--130},
  year={1976},
  publisher={Springer},
  doi={https://doi.org/10.1007/BF01608499}
}

@article{liu2016,
  title={Decoherence of topological qubit in linear and circular motions: decoherence impedance, anti-Unruh and information backflow},
  author={Liu, Pei-Hua and Lin, Feng-Li},
  journal={Journal of High Energy Physics},
  volume={2016},
  number={7},
  pages={1--35},
  year={2016},
  publisher={Springer},
doi={https://doi.org/10.1007/JHEP07(2016)084}
}

@article{maity2024,
doi = {10.1088/1751-8121/ad41a7},
url = {https://dx.doi.org/10.1088/1751-8121/ad41a7},
year = {2024},
month = {may},
publisher = {IOP Publishing},
volume = {57},
number = {21},
pages = {215302},
author = {Maity, Ananda G and Bhattacharya, Samyadeb},
journal = {Journal of Physics A: Mathematical and Theoretical}
}

@article{man2018,
  title = {Temperature effects on quantum non-Markovianity via collision models},
  author = {Man, Zhong-Xiao and Xia, Yun-Jie and Lo Franco, Rosario},
  journal = {Phys. Rev. A},
  volume = {97},
  issue = {6},
  pages = {062104},
  numpages = {10},
  year = {2018},
  month = {Jun},
  publisher = {American Physical Society},
  doi = {10.1103/PhysRevA.97.062104},
  url = {https://link.aps.org/doi/10.1103/PhysRevA.97.062104}
}

@article{mirkin2019,
  title = {Information backflow as a resource for entanglement},
  author = {Mirkin, Nicol\'as and Poggi, Pablo and Wisniacki, Diego},
  journal = {Phys. Rev. A},
  volume = {99},
  issue = {6},
  pages = {062327},
  numpages = {9},
  year = {2019},
  month = {Jun},
  publisher = {American Physical Society},
  doi = {10.1103/PhysRevA.99.062327},
  url = {https://link.aps.org/doi/10.1103/PhysRevA.99.062327}
}

@article{nielsen2000,
  title={Quantum information and quantum computation},
  author={Nielsen, Michael A and Chuang, Isaac L},
  journal={Cambridge: Cambridge University Press},
  volume={2},
  number={8},
  pages={23},
  year={2000}
}

@article{oi2003,
  title = {Interference of Quantum Channels},
  author = {Oi, Daniel K. L.},
  journal = {Phys. Rev. Lett.},
  volume = {91},
  issue = {6},
  pages = {067902},
  numpages = {4},
  year = {2003},
  month = {Aug},
  publisher = {American Physical Society},
  doi = {10.1103/PhysRevLett.91.067902},
  url = {https://link.aps.org/doi/10.1103/PhysRevLett.91.067902}
}

@article{paulson2022,
doi = {10.1088/1751-8121/acaadb},
url = {https://dx.doi.org/10.1088/1751-8121/acaadb},
year = {2022},
month = {dec},
publisher = {IOP Publishing},
volume = {55},
number = {50},
pages = {505302},
author = {Paulson, K G and Banerjee, Subhashish},
title = {Quantum speed limit time: role of coherence},
journal = {Journal of Physics A: Mathematical and Theoretical},
}

@article{poggi2017,
doi = {10.1209/0295-5075/118/20005},
url = {https://dx.doi.org/10.1209/0295-5075/118/20005},
year = {2017},
month = {jun},
publisher = {EDP Sciences, IOP Publishing and Società Italiana di Fisica},
volume = {118},
number = {2},
pages = {20005},
author = {Poggi, P. M. and Lombardo, F. C. and Wisniacki, D. A.},
title = {Driving-induced amplification of non-Markovianity in open quantum systems evolution},
journal = {Europhysics Letters}
}

@article{reich2015,
  title={Exploiting non-Markovianity for quantum control},
  author={Reich, Daniel M and Katz, Nadav and Koch, Christiane P},
  journal={Scientific reports},
  volume={5},
  number={1},
  pages={12430},
  year={2015},
  doi={https://doi.org/10.1038/srep12430},
  publisher={Nature Publishing Group UK London}
}

@article{rivas2010,
  title = {Entanglement and Non-Markovianity of Quantum Evolutions},
  author = {Rivas, \'Angel and Huelga, Susana F. and Plenio, Martin B.},
  journal = {Phys. Rev. Lett.},
  volume = {105},
  issue = {5},
  pages = {050403},
  numpages = {4},
  year = {2010},
  month = {Jul},
  publisher = {American Physical Society},
  doi = {10.1103/PhysRevLett.105.050403},
  url = {https://link.aps.org/doi/10.1103/PhysRevLett.105.050403}
}

@article{rubino2017,
author = {Giulia Rubino  and Lee A. Rozema  and Adrien Feix  and Mateus Araújo  and Jonas M. Zeuner  and Lorenzo M. Procopio  and Časlav Brukner  and Philip Walther },
title = {Experimental verification of an indefinite causal order},
journal = {Science Advances},
volume = {3},
number = {3},
pages = {e1602589},
year = {2017},
doi = {10.1126/sciadv.1602589},
URL = {https://www.science.org/doi/abs/10.1126/sciadv.1602589},
eprint = {https://www.science.org/doi/pdf/10.1126/sciadv.1602589}
}

@article{rubino2021,
  title = {Experimental quantum communication enhancement by superposing trajectories},
  author = {Rubino, Giulia and Rozema, Lee A. and Ebler, Daniel and Kristj\'ansson, Hl\'er and Salek, Sina and Allard Gu\'erin, Philippe and Abbott, Alastair A. and Branciard, Cyril and Brukner, \ifmmode \check{C}\else \v{C}\fi{}aslav and Chiribella, Giulio and Walther, Philip},
  journal = {Phys. Rev. Res.},
  volume = {3},
  issue = {1},
  pages = {013093},
  numpages = {19},
  year = {2021},
  month = {Jan},
  publisher = {American Physical Society},
  doi = {10.1103/PhysRevResearch.3.013093},
  url = {https://link.aps.org/doi/10.1103/PhysRevResearch.3.013093}
}

@article{sazim2021,
  title = {Classical communication with indefinite causal order for $N$ completely depolarizing channels},
  author = {Sazim, Sk and Sedlak, Michal and Singh, Kratveer and Pati, Arun Kumar},
  journal = {Phys. Rev. A},
  volume = {103},
  issue = {6},
  pages = {062610},
  numpages = {15},
  year = {2021},
  month = {Jun},
  publisher = {American Physical Society},
  doi = {10.1103/PhysRevA.103.062610},
  url = {https://link.aps.org/doi/10.1103/PhysRevA.103.062610}
}

@INPROCEEDINGS{shor1994,
  author={Shor, P.W.},
  booktitle={Proceedings 35th Annual Symposium on Foundations of Computer Science}, 
  title={Algorithms for quantum computation: discrete logarithms and factoring}, 
  year={1994},
  volume={},
  number={},
  pages={124-134},
  doi={10.1109/SFCS.1994.365700}
}

@article{simonov2022,
  title = {Work extraction from coherently activated maps via quantum switch},
  author = {Simonov, Kyrylo and Francica, Gianluca and Guarnieri, Giacomo and Paternostro, Mauro},
  journal = {Phys. Rev. A},
  volume = {105},
  issue = {3},
  pages = {032217},
  numpages = {12},
  year = {2022},
  month = {Mar},
  publisher = {American Physical Society},
  doi = {10.1103/PhysRevA.105.032217},
  url = {https://link.aps.org/doi/10.1103/PhysRevA.105.032217}
}

@article{simonov2025,
doi = {10.1088/1367-2630/ade5c4},
url = {https://dx.doi.org/10.1088/1367-2630/ade5c4},
year = {2025},
month = {jul},
publisher = {IOP Publishing},
volume = {27},
number = {7},
pages = {074502},
author = {Simonov, Kyrylo and Roy, Saptarshi and Guha, Tamal and Zimborás, Zoltán and Chiribella, Giulio},
title = {Activation of thermal states by coherently controlled thermalization processes},
journal = {New Journal of Physics}
}

@misc{tang2024,
      title={Demonstration of superior communication through thermodynamically free channels in an optical quantum switch}, 
      author={Hao Tang and Yu Guo and Xiao-Min Hu and Yun-Feng Huang and Bi-Heng Liu and Chuan-Feng Li and Guang-Can Guo},
      year={2024},
      eprint={2406.02236},
      archivePrefix={arXiv},
      primaryClass={quant-ph},
      url={https://arxiv.org/abs/2406.02236}, 
}

@article{thomas2108,
  title = {Thermodynamics of non-Markovian reservoirs and heat engines},
  author = {Thomas, George and Siddharth, Nana and Banerjee, Subhashish and Ghosh, Sibasish},
  journal = {Phys. Rev. E},
  volume = {97},
  issue = {6},
  pages = {062108},
  numpages = {8},
  year = {2018},
  month = {Jun},
  publisher = {American Physical Society},
  doi = {10.1103/PhysRevE.97.062108},
  url = {https://link.aps.org/doi/10.1103/PhysRevE.97.062108}
}

@article{thompson2018,
  title = {Causal Asymmetry in a Quantum World},
  author = {Thompson, Jayne and Garner, Andrew J. P. and Mahoney, John R. and Crutchfield, James P. and Vedral, Vlatko and Gu, Mile},
  journal = {Phys. Rev. X},
  volume = {8},
  issue = {3},
  pages = {031013},
  numpages = {15},
  year = {2018},
  month = {Jul},
  publisher = {American Physical Society},
  doi = {10.1103/PhysRevX.8.031013},
  url = {https://link.aps.org/doi/10.1103/PhysRevX.8.031013}
}

@article{utagi2024,
author = {Utagi, Shrikant and Banerjee, Subhashish and Srikanth, R.},
title = {On the eternal non-Markovianity of non-unital quantum channels},
journal = {International Journal of Quantum Information},
volume = {22},
number = {01},
pages = {2350039},
year = {2024},
doi = {10.1142/S0219749923500399},
URL = {https://doi.org/10.1142/S0219749923500399},
eprint = {https://doi.org/10.1142/S0219749923500399}
}

@article{vanrietvelde2021,
   title={Universal control of quantum processes using sector-preserving channels},
   volume={21},
   ISSN={1533-7146},
   url={http://dx.doi.org/10.26421/QIC21.15-16-5},
   DOI={10.26421/qic21.15-16-5},
   number={15 & 16},
   journal={Quantum Information and Computation},
   publisher={Rinton Press},
   author={Vanrietvelde, Augustin and Chiribella, Giulio},
   year={2021},
   month=nov, pages={1320–1352} }

@article{vaishy2022,
doi = {10.1088/1751-8121/ac677e},
url = {https://dx.doi.org/10.1088/1751-8121/ac677e},
year = {2022},
month = {may},
publisher = {IOP Publishing},
volume = {55},
number = {22},
pages = {225305},
author = {Vaishy, Ankit and Mitra, Subhadip and Bhattacharya, Samyadeb},
title = {Detecting genuine multipartite entanglement in three-qubit systems with eternal non-Markovianity},
journal = {Journal of Physics A: Mathematical and Theoretical},
}

@article{vilasini2024,
  title = {Embedding cyclic information-theoretic structures in acyclic space-times: No-go results for indefinite causality},
  author = {Vilasini, V. and Renner, Renato},
  journal = {Phys. Rev. A},
  volume = {110},
  issue = {2},
  pages = {022227},
  numpages = {49},
  year = {2024},
  month = {Aug},
  publisher = {American Physical Society},
  doi = {10.1103/PhysRevA.110.022227},
  url = {https://link.aps.org/doi/10.1103/PhysRevA.110.022227}
}

@article{vilasini2024a,
  title = {Fundamental Limits for Realizing Quantum Processes in Spacetime},
  author = {Vilasini, V. and Renner, Renato},
  journal = {Phys. Rev. Lett.},
  volume = {133},
  issue = {8},
  pages = {080201},
  numpages = {7},
  year = {2024},
  month = {Aug},
  publisher = {American Physical Society},
  doi = {10.1103/PhysRevLett.133.080201},
  url = {https://link.aps.org/doi/10.1103/PhysRevLett.133.080201}
}

@article{wolf2008,
  title = {Assessing Non-Markovian Quantum Dynamics},
  author = {Wolf, M. M. and Eisert, J. and Cubitt, T. S. and Cirac, J. I.},
  journal = {Phys. Rev. Lett.},
  volume = {101},
  issue = {15},
  pages = {150402},
  numpages = {4},
  year = {2008},
  month = {Oct},
  publisher = {American Physical Society},
  doi = {10.1103/PhysRevLett.101.150402},
  url = {https://link.aps.org/doi/10.1103/PhysRevLett.101.150402}
}

@article{xiang2014,
doi = {10.1209/0295-5075/107/54006},
url = {https://dx.doi.org/10.1209/0295-5075/107/54006},
year = {2014},
month = {sep},
publisher = {EDP Sciences, IOP Publishing and Società Italiana di Fisica},
volume = {107},
number = {5},
pages = {54006},
author = {Xiang, Guo-Yong and Hou, Zhi-Bo and Li, Chuan-Feng and Guo, Guang-Can and Breuer, Heinz-Peter and Laine, Elsi-Mari and Piilo, Jyrki},
title = {Entanglement distribution in optical fibers assisted by nonlocal memory effects},
journal = {Europhysics Letters}
}

@article{wu2020,
  title={Detecting non-Markovianity via quantified coherence: theory and experiments},
  author={Wu, Kang-Da and Hou, Zhibo and Xiang, Guo-Yong and Li, Chuan-Feng and Guo, Guang-Can and Dong, Daoyi and Nori, Franco},
  journal={npj Quantum Information},
  volume={6},
  number={1},
  pages={55},
  year={2020},
  publisher={Nature Publishing Group UK London},
doi={https://doi.org/10.1038/s41534-020-0283-3}
}

@article{zhao2020,
  title = {Quantum Metrology with Indefinite Causal Order},
  author = {Zhao, Xiaobin and Yang, Yuxiang and Chiribella, Giulio},
  journal = {Phys. Rev. Lett.},
  volume = {124},
  issue = {19},
  pages = {190503},
  numpages = {6},
  year = {2020},
  month = {May},
  publisher = {American Physical Society},
  doi = {10.1103/PhysRevLett.124.190503},
  url = {https://link.aps.org/doi/10.1103/PhysRevLett.124.190503}
}

\clearpage
\onecolumngrid   

\section{Appendix}

\subsection{Proof of Corollary \ref{cor_max_co_ps}}
According to the second condition of Proposition \ref{prop_max_co_ps}, we obtain the condition of activation of IB for the set of Bloch vectors given in Eq. (\ref{same_azi}) as follows.
\begin{align*}
    (\cos\theta_1-\cos\theta_2)^2>&\frac{M_{11}(\sin\theta_1\cos\phi-\sin\theta_2\cos\phi)^2+2M_{22}(\sin\theta_1\sin\phi-\sin\theta_2\sin\phi)^2}{3|M_{33}|}\\
    \implies(\cos\theta_1-\cos\theta_2)^2>&\frac{(\sin\theta_1-\sin\theta_2)^2(M_{11}\cos^2\phi+2M_{22}\sin^2\phi)}{3|M_{33}|}
\end{align*}
Since, the necessary condition of the activation is $t>\frac{1}{2}\ln 5,~|M_{33}|$ would take value as $e^{2t}-5$. Therefore, a straightforward calculation yields the above inequality as
\begin{align*}
    &\sin^2\left(\frac{\theta_1+\theta_2}{2}\right)>\frac{(3e^{2t}+1)\cos^2\phi+2(e^{2t}+3)\sin^2\phi}{3(e^{2t}-5)}\cos^2\left(\frac{\theta_1+\theta_2}{2}\right)\\
    \\
    \implies&2[1-\cos(\theta_1+\theta_2)]>\frac{5e^{2t}+7+(e^{2t}-5)\cos2\phi}{3(e^{2t}-5)}[1+\cos(\theta_1+\theta_2)]\\
    \\
    \implies&\cos(\theta_1+\theta_2)<\frac{e^{2t}-37-(e^{2t}-5)\cos2\phi}{11e^{2t}-23+(e^{2t}-5)\cos2\phi}
\end{align*}
This readily gives the range of $\theta_1+\theta_2$ as,
\begin{align*}
    \arccos\left[\frac{e^{2t}-37-(e^{2t}-5)\cos 2\Phi}{11e^{2t}-23+(e^{2t}-5)\cos 2\Phi}\right]<(\theta_{1}+\theta_{2})<2\pi-\arccos\left[\frac{e^{2t}-37-(e^{2t}-5)\cos 2\Phi}{11e^{2t}-23+(e^{2t}-5)\cos 2\Phi}\right]
\end{align*}

\subsection{Proof of Corollary \ref{cor_max_co_qs}}
Following the similar steps as in the above proof of Proposition \ref{cor_max_co_ps}, we obtain the condition of activation of IB under the quantum-switch setup for those states as follows.
\begin{align*}
    (\cos\theta_1-\cos\theta_2)^2>&\frac{N_{11}(\sin\theta_1-\sin\theta_2)^2}{2|N_{33}|}
\end{align*}
Here, the necessary condition of the activation is $t>\frac{1}{2}\ln (1+2\sqrt{2})$, hence $|N_{33}|=\frac{e^{4t}-2e^{2t}-7}{7e^{4t}+2e^{2t}-1}$. Therefore, the above inequality turns into the following simplified form.
\begin{align*}
    &\sin^2\left(\frac{\theta_1+\theta_2}{2}\right)>\frac{3e^{4t}+2e^{2t}+3}{2(e^{4t}-2e^{2t}-7)}\cos^2\left(\frac{\theta_1+\theta_2}{2}\right)\\
    \\
    \implies&-\frac{5e^{4t}-2e^{2t}-11}{2(e^{4t}-2e^{2t}-7)}\cos(\theta_1+\theta_2)>\frac{e^{4t}+6e^{2t}+17}{2(e^{4t}-2e^{2t}-7)}\\
    \\
    \implies&\cos(\theta_1+\theta_2)<-\frac{e^{4t}+6e^{2t}+17}{5e^{4t}-2e^{2t}-11}
\end{align*}
Thus we have the range of $\theta_1+\theta_2$ in this configuration as,
\begin{align*}
    \pi-\arccos\left[\frac{e^{4t}+6e^{2t}+17}{5e^{4t}-2e^{2t}-11}\right]<(\theta_{1}+\theta_{2})<\pi+\arccos\left[\frac{e^{4t}+6e^{2t}+17}{5e^{4t}-2e^{2t}-11}\right]
\end{align*}

\subsection{Proof of Corollary \ref{cor_rc_ps}}

Here, we restrict our discussion on the $xz$-plane. Hence, there can not be any contribution of $(r_2-s_2)^2$. Therefore, the second condition of Proposition \ref{prop_rc_ps} suggests,
\begin{align*}
       (r_3-s_3)^2>&\frac{M(p)_{11}(r_1-s_1)^2}{(2+p)|M(p)_{33}|} \\   
       \\
    \implies  (\cos\theta_1-\cos\theta_2)^2>&\frac{M(p)_{11}(\sin\theta_1-\sin\theta_2)^2}{(2+p)|M(p)_{33}|}
\end{align*}
Following the same arguments as the previous proofs, the time scale we have to consider here as $t>\frac{1}{2}\ln\left(1+\frac{4}{p}\right)$ as provided by Proposition \ref{prop_rc_ps}. So that, $|M(p)_{33}|=-M(p)_{33}$. Therefore, the above inequality becomes,
\begin{align*}
    \sin^2\left(\frac{\theta_1+\theta_2}{2}\right) & >  \frac{M(p)_{11}}{(2+p)|M(p)_{33}|}\cos^2\left(\frac{\theta_1+\theta_2}{2}\right)\\
    \\
    \implies  \cos(\theta_1+\theta_2) & <  \frac{(2+p)M(p)_{33}-M(p)_{11}}{(2+p)M(p)_{33}+M(p)_{11}}\\
    &=-\frac{(2-p-p^2)e^{2t}+(10+5p+p^2)}{(2+p)(1+p)e^{2t}-(6+p)(1+p)}
\end{align*}
Hence, we obtain the range of $(\theta_1+\theta_2)$ as 
\begin{align*}
    \pi-\arccos\left[\frac{(2-p-p^2)e^{2t}+(10+5p+p^2)}{(2+p)(1+p)e^{2t}-(6+p)(1+p)}\right]<(\theta_1+\theta_2)<\pi+\arccos\left[\frac{(2-p-p^2)e^{2t}+(10+5p+p^2)}{(2+p)(1+p)e^{2t}-(6+p)(1+p)}\right]
\end{align*}

\subsection{Proof of Corollary \ref{cor_rc_qs}}

Since, there is no contribution of $(r_2-s_2)^2$, we get from Proposition \ref{prop_rc_qs}
\begin{align*}
   (r_3-s_3)^2>&\frac{f_1(p,t)N(p)_{11}(r_1-s_1)^2}{2f_2(p,t)|N(p)_{33}|} \\   
       \\
    \implies  (\cos\theta_1-\cos\theta_2)^2>&\frac{f_1(p,t)N(p)_{11}(\sin\theta_1-\sin\theta_2)^2}{2f_2(p,t)|N(p)_{33}|} 
\end{align*}
Since, Proposition \ref{prop_rc_qs} suggested the time regime for this case $t>\frac{1}{2}\ln\left(1+2\sqrt{1+\frac{1}{p}}\right)$, we will consider $|N(p)_{33}|=-N(p)_{33}$ and the above inequality reduces to
\begin{align*}
   \sin^2\left(\frac{\theta_1+\theta_2}{2}\right) & >  \frac{f_1(p,t)N(p)_{11}}{2f_2(p,t)|N(p)_{33}|}\cos^2\left(\frac{\theta_1+\theta_2}{2}\right)\\
    \\ 
    \implies  \cos(\theta_1+\theta_2) & <  \frac{2f_2(p,t)N(p)_{33}-f_1(p,t)N(p)_{11}}{2f_2(p,t)N(p)_{33}+f_1(p,t)N(p)_{11}}
\end{align*}
Substituting the values of $f_1(p,t),~f_2(p,t),~N(p)_{11}$ and $N(p)_{33}$ from the main manuscript, the above bound on the state parameter reduces to,
Hence, we obtain the permissible range of $(\theta_1+\theta_2)$ as 
\begin{align*}
    \cos(\theta_1+\theta_2) & <-\frac{\sum_{j=0}^4A_je^{2jt}}{A_5}\\ 
    \implies 
    \pi-\arccos&\left(\frac{\sum_{j=0}^4A_je^{2jt}}{A_5}\right)<(\theta_1+\theta_2)<\pi+\arccos\left(\frac{\sum_{j=0}^4A_je^{2jt}}{A_5}\right),
\end{align*}
where, $\{A_j\}_{j=0}^5$ are given in the main paper.

\end{document}